\documentclass[a4paper,10pt]{article}
\usepackage{amsmath}
\usepackage{amssymb}
\usepackage{lmodern}

\usepackage{sectsty}
\sectionfont{\fontsize{12}{15}\selectfont}

\usepackage{extsizes}
\usepackage{textgreek}
\usepackage{multicol}
\usepackage{fullpage}
\usepackage{setspace}
\usepackage{pdflscape}
\usepackage[export]{adjustbox}
\usepackage{chngcntr}
\usepackage{pdfpages}
\usepackage{subcaption}

\usepackage{sansmath}
\usepackage[a4paper,margin=1in,headsep=7.5mm,footnotesep=2.2\baselineskip]{geometry}
\usepackage{times}
\usepackage{newtxtext,newtxmath}
\usepackage{framed}
\usepackage{relsize}
\usepackage{mathtools}
\usepackage{xfrac}
\usepackage{graphicx}
\usepackage{tikz}
\usepackage[Symbol]{upgreek}
\usepackage{array}
\usepackage{accents}
\usepackage[T1]{fontenc}
\usepackage{esint}
\usepackage{lipsum}
\usepackage{float}
\usepackage{epstopdf}
\usepackage{colortbl,xcolor}
\usepackage[font=small,labelfont=bf]{caption}
\usepackage{enumitem}
\usepackage{fancyhdr}
\usepackage{mathtools}
\usepackage{changepage}

\allowdisplaybreaks

\usepackage[hang,flushmargin,bottom]{footmisc}

\usepackage[numbers]{natbib}
\usepackage[colorlinks=true,
linkcolor=blue,
urlcolor=blue,
citecolor=red,
hypertexnames=false]{hyperref}

\newcommand{\ccos}{\mathsf{Cos}}

\usepackage{rotating}

\makeatletter
\DeclareSymbolFont{largesymbolsB}{U}{esint}{m}{n}
\re@DeclareMathSymbol{\intop}{\mathop}{largesymbolsB}{'001}
\def\int{\intop\nolimits}
\makeatother

\newcolumntype{L}[1]{>{\raggedright\let\newline\\\arraybackslash\hspace{0pt}}m{#1}}
\newcolumntype{C}[1]{>{\centering\let\newline\\\arraybackslash\hspace{0pt}}m{#1}}
\newcolumntype{R}[1]{>{\raggedleft\let\newline\\\arraybackslash\hspace{0pt}}m{#1}}

\newcommand{\qag}[1]{{\fontfamily{qag}\selectfont #1}}
\newcommand{\tit}[1]{{\fontfamily{ppl}\selectfont \textit{#1}}}

\definecolor{armygreen}{rgb}{0.14, 0.71, 0.15}
\definecolor{darkgreen}{rgb}{0.08, 0.48, 0.18}
\definecolor{darkred}{rgb}{0.86, 0.153, 0.153}
\definecolor{azure}{rgb}{0.0, 0.5, 1.0}
\definecolor{bole}{rgb}{0.82, 0.57, 0.22}
            
\newcommand{\fdotlo}{${-2.93\times 10^{-9}\,\mathrm{Hz/s}}$ } 
\newcommand{\fdothi}{${5.53\times 10^{-10}\,\mathrm{Hz/s}}$ } 
\newcommand{\WUcputimeHours}{${6\,\mathrm{hours}}$} 
 
\newcommand{\totaltemplates}{${5.6\times 10^{16}}$ } 
\def\EatH{Einstein@Home }

\begin{document}
\newpage
\topskip15pt
\fancyhead[L]{\footnotesize\tit{Avneet Singh et al}}
\fancyhead[R]{{\footnotesize \tit{published as} \href{https://journals.aps.org/prd/abstract/10.1103/PhysRevD.94.064061}{\tit{\textbf{Physical Review D}} 94(6):064061}}}
\begin{flushleft}
\textbf{\large Results of an all-sky high-frequency  \EatH search for continuous gravitational waves in LIGO's fifth science run}\linebreak\linebreak
{\small Avneet Singh$^\mathrm{1,2,3,a}$\let\thefootnote\relax\footnote{$^\mathrm{a}$ avneet.singh@aei.mpg.de}, Maria Alessandra Papa$^\mathrm{1,2,4,b}$\let\thefootnote\relax\footnote{$^\mathrm{b}$ maria.alessandra.papa@aei.mpg.de}, Heinz-Bernd Eggenstein$^\mathrm{2,3}$, Sylvia Zhu$^\mathrm{1,2}$, Holger Pletsch$^\mathrm{2,3}$, Bruce Allen$^\mathrm{2,4,3}$, Oliver Bock$^\mathrm{2,3}$, Bernd Maschenchalk$^\mathrm{2,3}$, Reinhard Prix$^\mathrm{2,3}$, Xavier Siemens$^\mathrm{4}$
}\linebreak\linebreak
{\footnotesize $^1$ Max-Planck-Institut f{\"u}r Gravitationsphysik, am M{\"u}hlenberg 1, 14476, Potsdam-Golm\\
$^2$ Max-Planck-Institut f{\"u}r Gravitationsphysik, Callinstra{$\upbeta$}e 38, 30167, Hannover\\
$^3$ Leibniz Universit{\"a}t Hannover, Welfengarten 1, 30167, Hannover\\
$^4$ University of Wisconsin-Milwaukee, Milwaukee, Wisconsin 53201, USA
}\linebreak\linebreak
\setcounter{footnote}{0}
\vspace{-0.3in}
\end{flushleft}
\begin{center}
\begin{abstract}
We present results of a high-frequency all-sky search for continuous gravitational waves from isolated compact objects in LIGO's 5th Science Run ($\mathsf{S5}$) data, using the computing power of the \EatH volunteer computing project. This is the only dedicated continuous gravitational wave search that probes this high frequency range on $\mathsf{S5}$ data. We find no significant candidate signal, so we set 90\%-confidence level upper-limits on continuous gravitational wave strain amplitudes. At the lower end of the search frequency range, around 1250$\,$Hz, the most constraining upper-limit is $5.0\times 10^{-24}$, while at the higher end, around 1500$\,$Hz, it is $6.2\times 10^{-24}$. Based on these upper-limits, and assuming a fiducial value of the {\text{principal moment of inertia}} of $10^{38}$kg$\,$m$^2$, we can exclude objects with ellipticities higher than roughly $2.8\times10^{-7}$ within 100$\,$pc of Earth with rotation periods between $1.3$ and $1.6$ milliseconds.
\end{abstract}
\end{center}
\begin{multicols}{2}
\section{Introduction}
\label{section:intro}
Ground-based gravitational wave (GW) detectors will be able to detect a continuous gravitational wave signal from a spinning deformed compact object provided that it is spinning with a rotational period between roughly $1$ and $100$ milliseconds, that it is sufficiently close to Earth and it is sufficiently ``bumpy''. Blind searches for continuous gravitational waves probe the whole sky and broad frequency ranges, looking for this type of objects. 

In this paper, we present the results of an all-sky \EatH search for continuous, nearly monochromatic, high-frequency gravitational waves in data from LIGO's 5th Science Run ($\mathsf{S5}$). A number of searches have been carried out on LIGO data \citep{S6BucketStage0,EarlyS5Paper,FullS5Semicoherent,S5EHHough,S4IncoherentPaper,S2FstatPaper} targeting lower frequency ranges. The only other search covering frequencies up to 1500$\,$Hz was conducted on $\mathsf{S6}$ data \citep{S6PowerFlux} taken at least 3 years apart from the data used here. Our search results are only 33\% less sensitive than those of \citet{S6PowerFlux}, even though the $\mathsf{S5}$ data is less sensitive than the $\mathsf{S6}$ data by more than a factor of 2. The search method presented here anticipates the procedure that will be used on the advanced detector (aLIGO) data.

This search can be considered an extension of the $\mathsf{S5}$ \EatH search \cite{S5EHHough} although it employs a different search technique: this search uses the {\fontfamily{ppl}\selectfont\textit{Global Correlation Transform}} (GCT) method to combine results from coherent $\mathcal{F}${\fontfamily{ppl}\selectfont\textit{-statistic}} searches  \citep{Holger2008,Holger2010}, as opposed to the previous \EatH search \cite{S5EHHough} that employed the {\fontfamily{ppl}\selectfont\textit{Hough-transform}} method to perform this combination. In the end, at fixed computing resources, these two search methods are comparable in sensitivity. However, a semi-coherent $\mathcal{F}${\fontfamily{ppl}\selectfont\textit{-statistic}} search is more efficient when considering a broad spin-down range, and for the \EatH searches we have decided to adopt it as our ``work horse''.

We do not find any significant signal(s) among the set of searched waveforms. Thus, we set 90\%-confidence upper-limits on continuous gravitational wave strain amplitudes; near the lower end of the search frequency range between 1253.217--1255.217$\,$Hz, the most constraining upper-limit is $5.0\times 10^{-24}$, while toward the higher end of the search frequency range nearing 1500$\,$Hz, the upper-limit value is roughly $6.2\times 10^{-24}$. Based on these upper-limits, we can exclude certain combinations of signal frequency, star deformation (ellipticity) and distance values. We show with this search that even with $\mathsf{S5}$ data from the first generation of GW detectors, such constraints do probe interesting regions of source parameter space.

\section{The Data}
\label{sec:S6intro} 
The LIGO gravitational wave network consists of two detectors, H1 in Hanford (Washington) and L1 in Livingston (Louisiana), separated by a 3000-km baseline. 
The $\mathsf{S5}$ run lasted roughly two years between GPS time 815155213 sec (Fri, Nov 04, 16:00:00 UTC 2005) and 875145614 sec (Sun, Sep 30, 00:00:00 UTC 2007). This search uses data spanning this observation period, and during this time, H1 and L1 had duty-factors of 78\% and 66\% respectively \citep{Duty1,Duty2}. The gaps in this data-set are due to environmental or instrumental disturbances, or scheduled maintenance periods.

We follow \cite{S5EHHough, S2FstatPaper}, where the calibrated and high-pass filtered data from each detector is partitioned in 30-minute chunks and each chunk is Fourier-transformed after the application of a steep {{Tukey window}}. The set of \textbf{S}hort (time-baseline) \textbf{F}ourier \textbf{T}ransforms ({\fontfamily{ppl}\selectfont\textit{abbrev}}. $\mathsf{SFT}$) that ensues, is the input data for our search.

We further follow \citep{S5EHHough}, where frequency bands known to contain spectral disturbances have been removed from the analysis. In fact, such data has been substituted with fake Gaussian noise at the same level as the neighboring undisturbed data; in table \ref{table:lines}, we list these bands.
\section{The Search}
\label{sec:Search} 

The search presented here is similar to the search on $\mathsf{S6}$ data, reported in \citep{S6BucketStage0}. Our reference target signal is given by (1)-(4) in \citep{EarlyS5Paper}; at emission, the signal is nearly monochromatic, typically with a small `spin-down'. The signal waveform in the detector data is modulated in frequency because of the relative motion between the compact object and the detector; a modulation in amplitude also occurs because of the variation of the sensitivity of the detector with time across the sky. 

The most sensitive search technique that one could use is a {{fully-coherent combination}} of the detectors' data, matched to the waveform that one is looking for. The (amplitude) sensitivity of such a method increases with the square-root of the time-span of the data used. However, the computational cost to resolve different waveforms increases very rapidly with increasing time-span of the data, and this makes a {{fully-coherent search}} over a large frequency range  computationally unfeasible when using months of data. This is the main reason why {\fontfamily{ppl}\selectfont\textit{semi-coherent search}} methods have been developed. These methods perform coherent searches over shorter stretches of data, called {\fontfamily{ppl}\selectfont\textit{segments}}, and then combine the results with incoherent techniques.

This search covers waveforms from the entire sky, with frequencies in a 250$\,$Hz range from 1249.717$\,$Hz to 1499.717$\,$Hz, and with a first-order spin-down between \fdotlo\! and \fdothi\!\hspace{-0.015in}, similar to previous \EatH searches. We use a ``stack--slide'' {{semi-coherent search}} procedure implemented with the GCT method \citep{Holger2008, Holger2010}. 
\end{multicols}
\begin{figure}[H]
\centering\includegraphics[width=158mm]{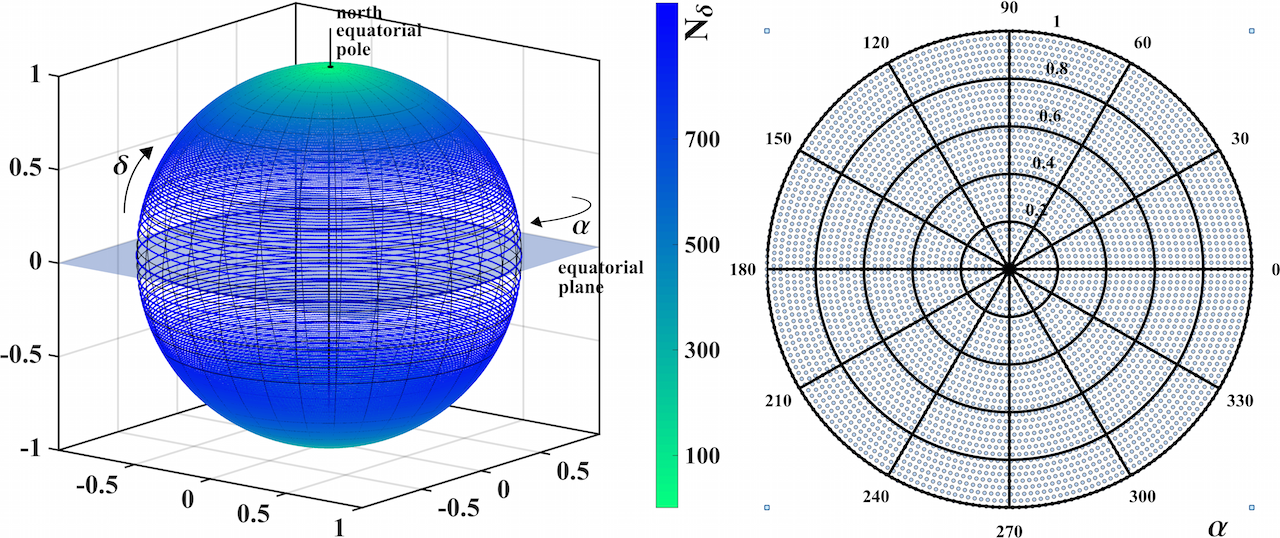}
\caption{{\small Tilling of sky-grid for the frequency band 1240-1250$\,$Hz; $d_\mathsf{sky} = 6.6\times 10^{-4}$ for this band. In the left panel, we show the sky-grid points on the celestial sphere; the color-code traces the number of sky-grid points, $\mathrm{N}_\delta$, as a function of equatorial latitude $\delta$. The right panel is a polar plot of the northern equatorial hemisphere of the same sky-grid but with density scaled down by a factor of 4 to allow for better viewing. In the polar plot, $\uptheta=\alpha$ and $r=\ccos(\delta)$.}}
\label{fig:sky}
\end{figure}
\begin{multicols}{2}
{\noindent}The data is divided into $\mathrm{N}_\mathsf{seg}$ segments, each spanning $\mathrm{T}_\mathsf{coh}$ in time. The coherent multi-detector $\mathcal{F}$-statistic \cite{FStatSchutz} is computed on each segment for all the points on a coarse $\uplambda_\mathrm{c}\equiv\{f_\mathrm{c},\dot{f}_\mathrm{c},\alpha_\mathrm{c},\delta_\mathrm{c}\}$ signal waveform parameter grid, and then results from the individual segments are summed, one per segment, to yield the final core detection-statistic $\overline{\mathcal{F}}$, as shown in \eqref{eq:fstat}; $\alpha, \delta$ are the equatorial sky coordinates of the source position, while $f$ and $\dot{f}$ are the frequency and first-order spin-down of the signal respectively. Depending on which $\uplambda_\mathrm{c}$ parameter points are taken on the coarse grid for each segment in this sum, the result will approximate the detection-statistic computed on a $\uplambda_{f}$ parameter point on a finer grid: 
\begin{equation}
\overline{\mathcal{F}}(\uplambda_{f}):= \frac{1}{\mathrm{N}_\mathsf{seg}}\sum_{i=1}^{\mathrm{N}_\mathsf{seg}}\mathcal{F}(\uplambda^i_\mathrm{c}).
\label{eq:fstat}
\end{equation}  
In a ``stack--slide'' search in Gaussian noise, $\mathrm{N}_\mathsf{seg}\times2\overline{\mathcal{F}}$ follows a $\upchi^2_{4\mathrm{N}_\mathsf{seg}}$ chi-squared distribution with ${4\mathrm{N}_\mathsf{seg}}$ degrees of freedom. 

The most important search parameters are then: $\mathrm{N}_\mathsf{seg}$, $\mathrm{T}_\mathsf{coh}$, the signal parameter search grids  $\uplambda_\mathrm{c},\,\uplambda_{f}$, the total spanned observation time $\mathrm{T}_\mathsf{obs}$, and finally the ranking statistic used to rank parameter space cells i.e. $2\overline{\mathcal{F}}$.  

The grid-spacing in frequency $\delta\hspace{-0.015in}f$ and spin-down $\delta\hspace{-0.015in}\dot{f}$ are constant over the search range. The same frequency spacing and sky grid is used for the coherent analysis and in the incoherent summing. The spin-down spacing of the incoherent analysis is finer by a factor of $\gamma$ with respect to that of the coherent analysis. In table \ref{table:parameters}, we summarize the search parameters.

\begin{figure}[H]
\centering\includegraphics[width=78mm]{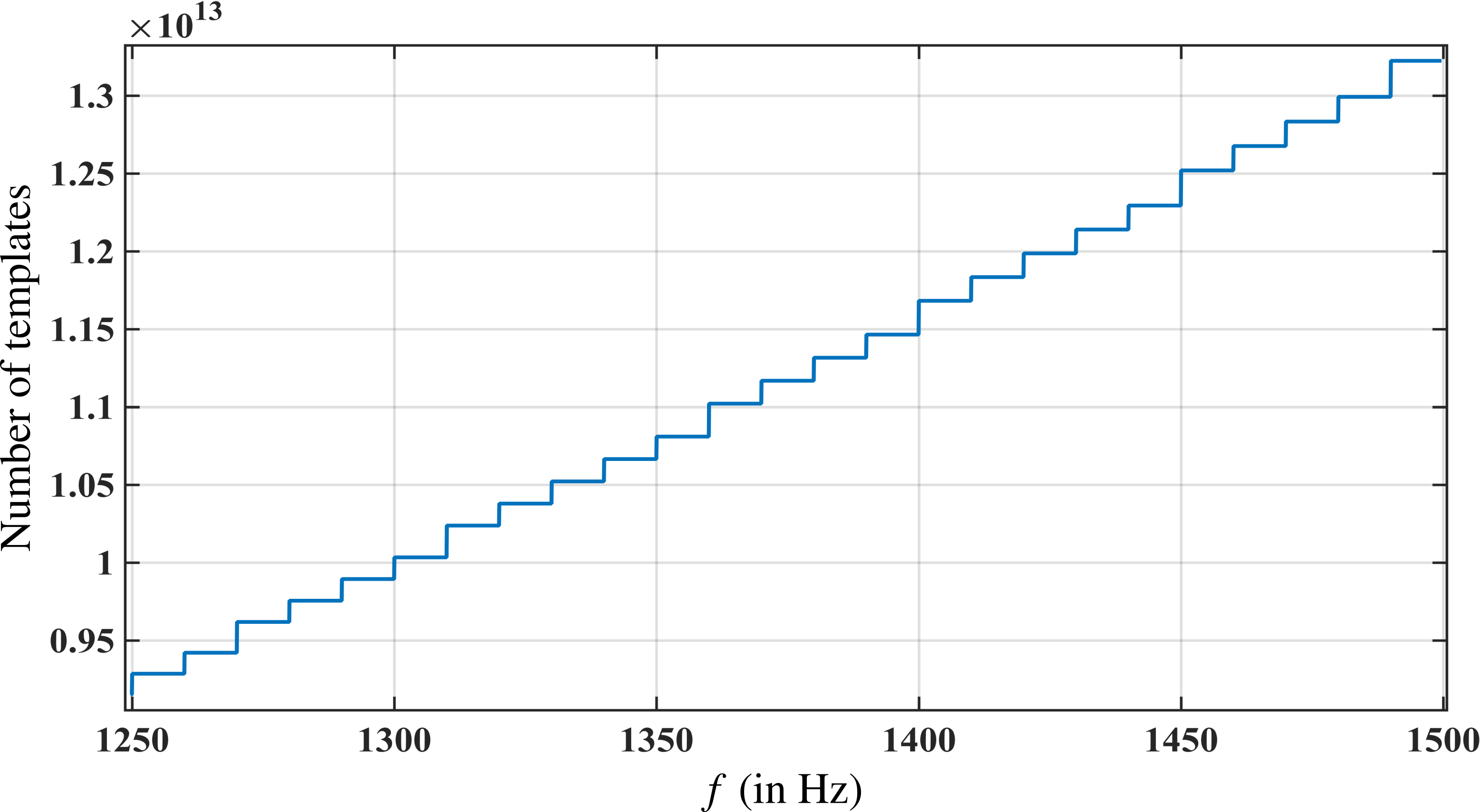}
\caption{{Number of templates searched in 50$\,$mHz bands. The variation in template count arises from the variation is number of sky-grid points every 10$\,$Hz in frequency. Each 50$\,$mHz band contributes roughly $6.3\times10^{7}$ templates in frequency and spin-down (on the finer grid refined by {\fontfamily{ppl}\selectfont \textit{refinement factor}} $\gamma$.)}}
\label{fig:templates}
\end{figure}

The sky-grid for the search is constructed by tiling the projected equatorial plane uniformly with squares of edge length $d_\mathsf{sky}$. The length of the edge of the squares is a function of the frequency $f$ of the signal, and parameterized in terms of a so-called {\fontfamily{ppl}\selectfont\textit{sky-mismatch parameter}} ($\mathrm{m}_\mathsf{sky}$) as
\begin{equation}
d_\mathsf{sky}=\frac{1}{f}\frac{\sqrt{\mathrm{m}_\mathsf{sky}}}{\uppi\tau_\mathsf{E}},
\label{eq:skygrid}
\end{equation}  
where, $\tau_\mathsf{E} = 0.021\,\mathrm{seconds}$ and $\mathrm{m}_\mathsf{sky}$ = 0.3, also given in table \ref{table:parameters}. The sky-grids are constant over 10$\,$Hz-wide frequency bands, and are calculated for the highest frequency in the band. In figure \ref{fig:sky}, we illustrate an example of the sky-grid. The total number of templates in 50$\,$mHz bands as a function of frequency is shown in figure \ref{fig:templates}. This search explores a total of \totaltemplates\! waveform templates across the $\uplambda_{f}\equiv\{f_f,\dot{f}_f,\alpha_f,\delta_f\}$ parameter space.

The search is divided into work-units ({\fontfamily{ppl}\selectfont\textit{abbrev}}. $\mathrm{WU}$), each searching a very small sub-set of template waveforms. The WU are sent to \EatH volunteers and each WU occupies the volunteer/host computer for roughly \WUcputimeHours. One such WU covers a 50$\,$mHz band, the entire spin-down range, and 139--140 points in the sky.  6.4 million different WU are necessary to cover the whole parameter space. Each WU returns a ranked list of the most significant $10^4$ candidates found in the parameter space that it searched.\vspace{-5pt} 
\section{Identification of Undisturbed Bands}
\label{sec:undisturbedbands} 
In table \ref{table:lines}, we list the central frequencies and bandwidths of $\mathsf{SFT}$ data known to contain spectral lines from instrumental artefacts. These frequency regions were identified before the \EatH run, and we were able to replace the corresponding data with Gaussian noise matching the noise level of neighbouring quiet bands. Consequently, some search results have contributions from this `fake data'. The intervals in signal-frequency where the search results come entirely from fake data are indicated as {\fontfamily{ppl}\selectfont \textit{All Fake Data}} in table \ref{table:ranges}. In these intervals of signal-frequency, we effectively do not have search results. The other three columns in table  \ref{table:ranges} provide signal-frequency intervals where results {\fontfamily{ppl}\selectfont \textit{might}} have contributions from fake data. In these regions, depending on the signal parameters, the detection efficiency might be affected.

Despite the removal of known disturbances from the data, it still contains unknown noise artefacts producing $2\overline{\mathcal{F}}$ values that do not follow the expected distribution for Gaussian noise. These artifacts usually have narrow-band characteristics; we identify such `disturbed' signal-frequency intervals in the search results and exclude them from further post-processing analysis. \vspace{-5pt}
\begin{table}[H]
\begin{center}
\bgroup
\def\arraystretch{1.2}
\begin{tabular}{|l|l|}
\hline
\hline
\textbf{Quantity} & \textbf{Value}\\
\hline
\hline
$\mathrm{T}_\mathsf{coh}$ (hours) & 30.0\\
\hline
$\mathrm{T}_\mathsf{obs}$ (days) & 653.18\\
\hline
${t}_\mathsf{ref}$ (GPS seconds) & 847063082.5\\
\hline
$\mathrm{N}_\mathsf{seg}$ & 205\\
\hline
$\delta\hspace{-0.015in}f_\mathrm{c}$ (Hz) & $6.71\times10^{-6}$\\
\hline
$\delta\hspace{-0.015in}\dot{f}_\mathrm{c}$ (Hz/s) & $5.78\times10^{-10}$\\
\hline
$\gamma$ & 1399\\
\hline
$\mathrm{m}_\mathsf{sky}$ & 0.30\\
\hline
\hline
\end{tabular}
\egroup
\end{center}
\caption{Search parameters for the search. ${t}_\mathsf{ref}$ is the reference time that defines the frequency and spin-down values.}
\label{table:parameters}
\end{table}

\vspace{-0.3in}
\begin{figure}[H]
\centering\includegraphics[width=78mm]{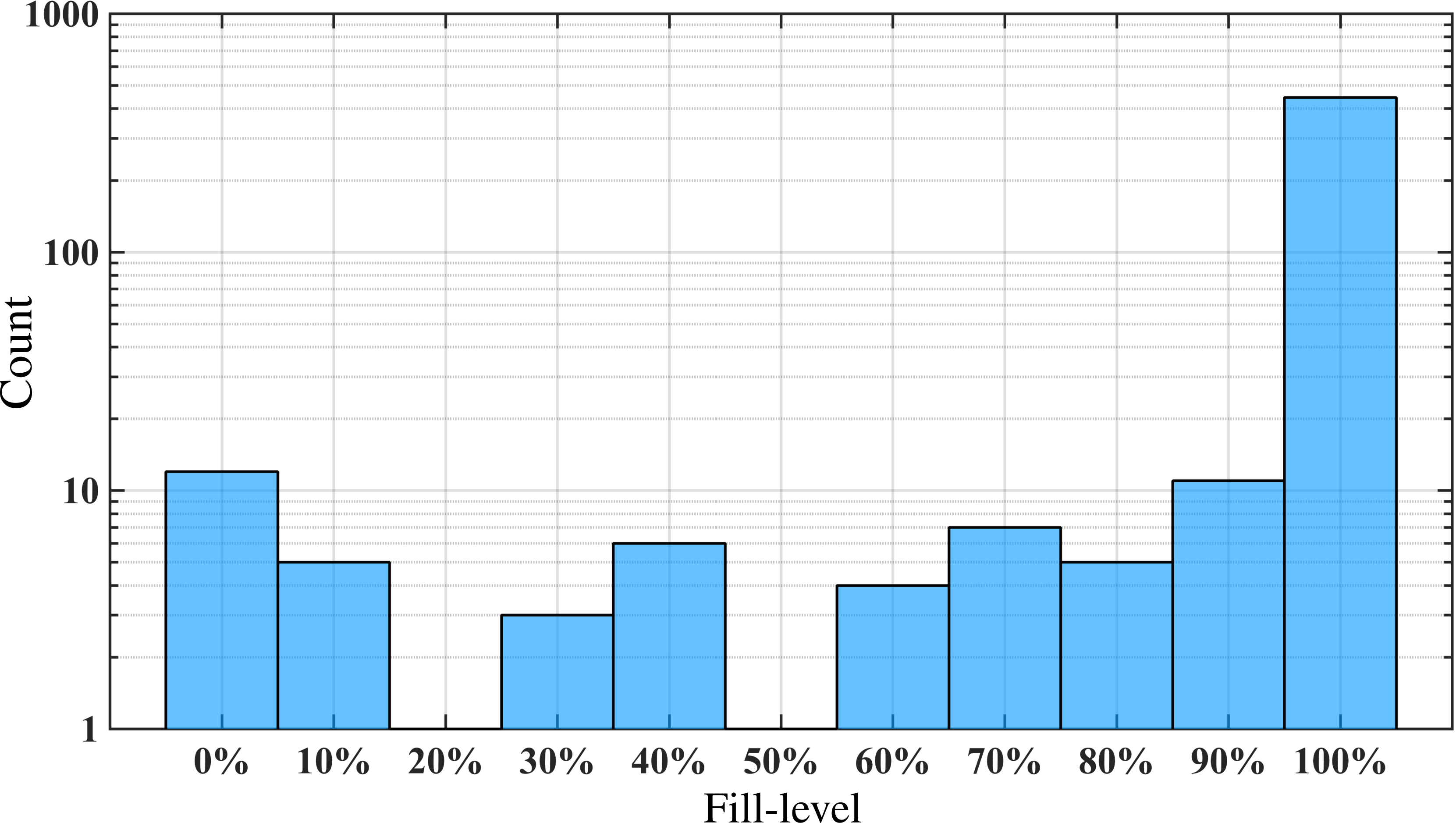}
\caption{{\small Distribution of fill-levels of 0.5$\,$Hz bands.}}
\label{fig:fill}
\end{figure}

\vspace{-0.2in}
{\noindent}The benefit of such exclusions is that, in the remaining `undisturbed' bands, we can rely on semi-analytic predictions for the significance of the observed $2\overline{\mathcal{F}}$ values, and we can set a uniform detection criterion across the entire parameter space. It is true that we forego the possibility of detecting a target signal in the `disturbed' frequency intervals. However, to perform reliable analysis in these intervals, ad-hoc studies and tuning of the procedures would need to be set up. These additional procedures would require as much, if not longer, than the time spent on the `undisturbed' data set. Moreover, since the `undisturbed' intervals in data comprise over 95\% of the total data, we believe this is a reasonable choice. In the future, a focused effort on the analysis of `disturbed bands' could attempt to recover some sensitivity in those regions.
The identification of undisturbed bands is carried out via a {\fontfamily{ppl}\selectfont\textit{visual inspection method}}. This visual inspection of the data is performed by two scientists who look at various distributions of the $2\overline{\mathcal{F}}$ values in the $\{f,\dot{f}\}$ parameter space in 50$\,$mHz bands. They rank these 50$\,$mHz bands with 4 numbers: 0,1,2,3; a `0'
\begin{figure}[H]
\centering\includegraphics[width=75mm]{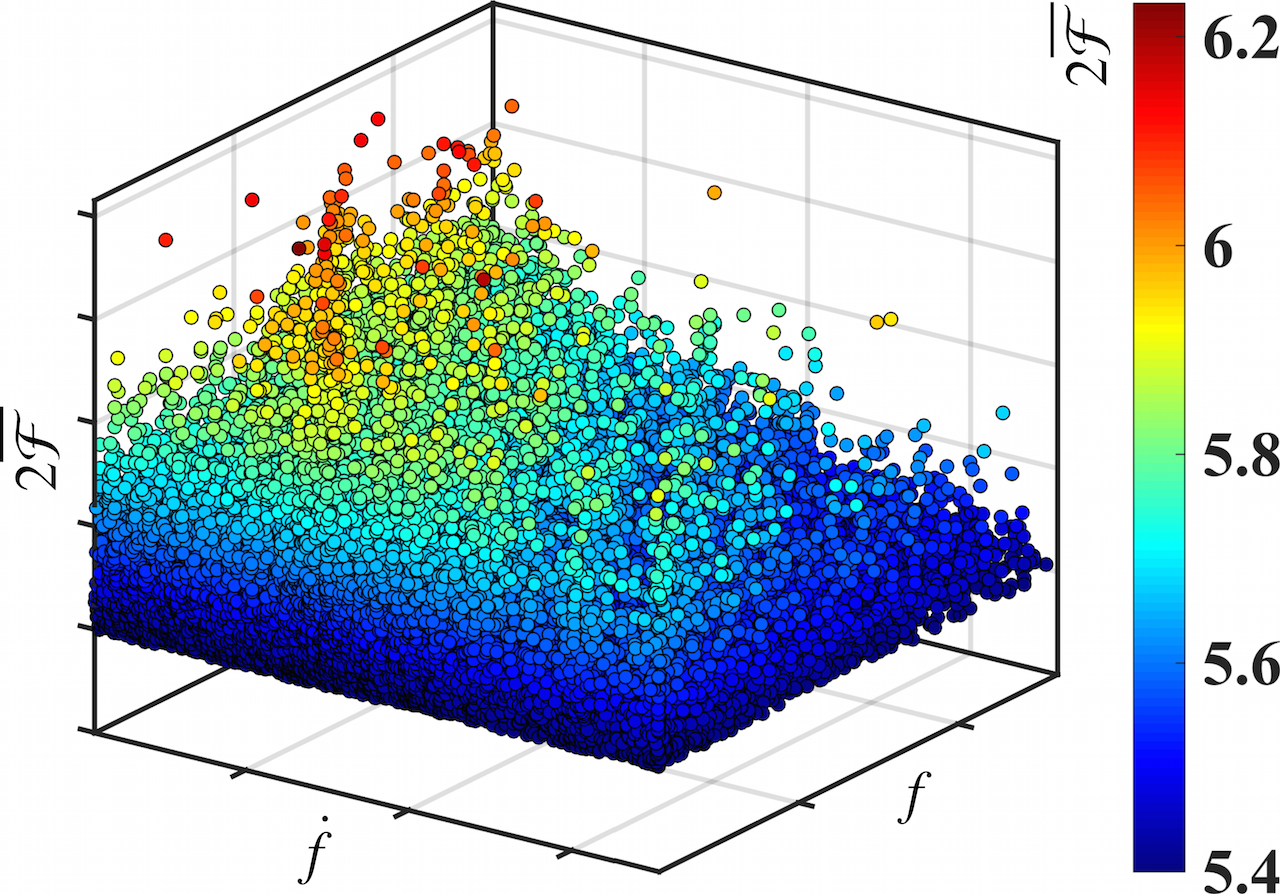}
\centering\includegraphics[width=75mm]{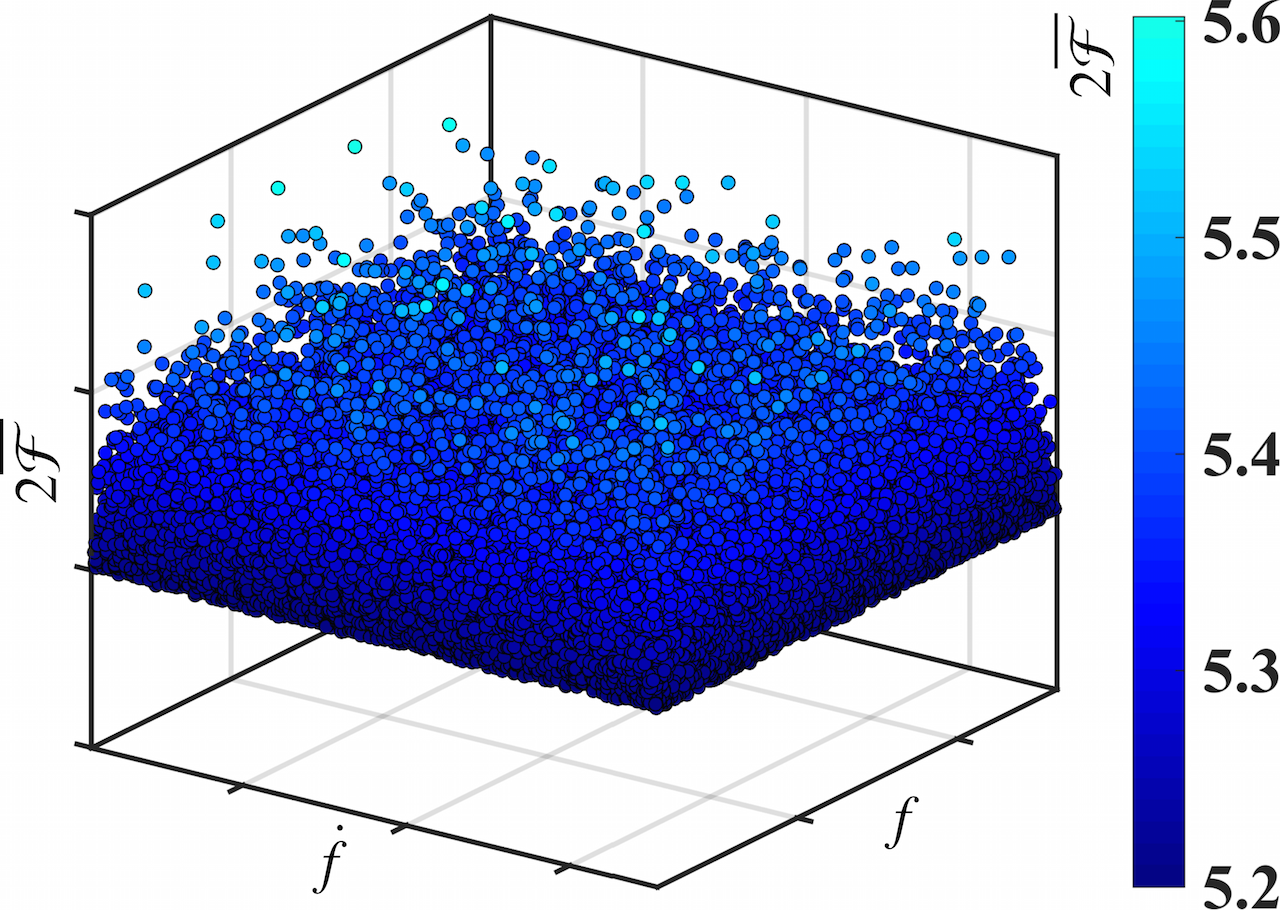}
\centering\includegraphics[width=75mm]{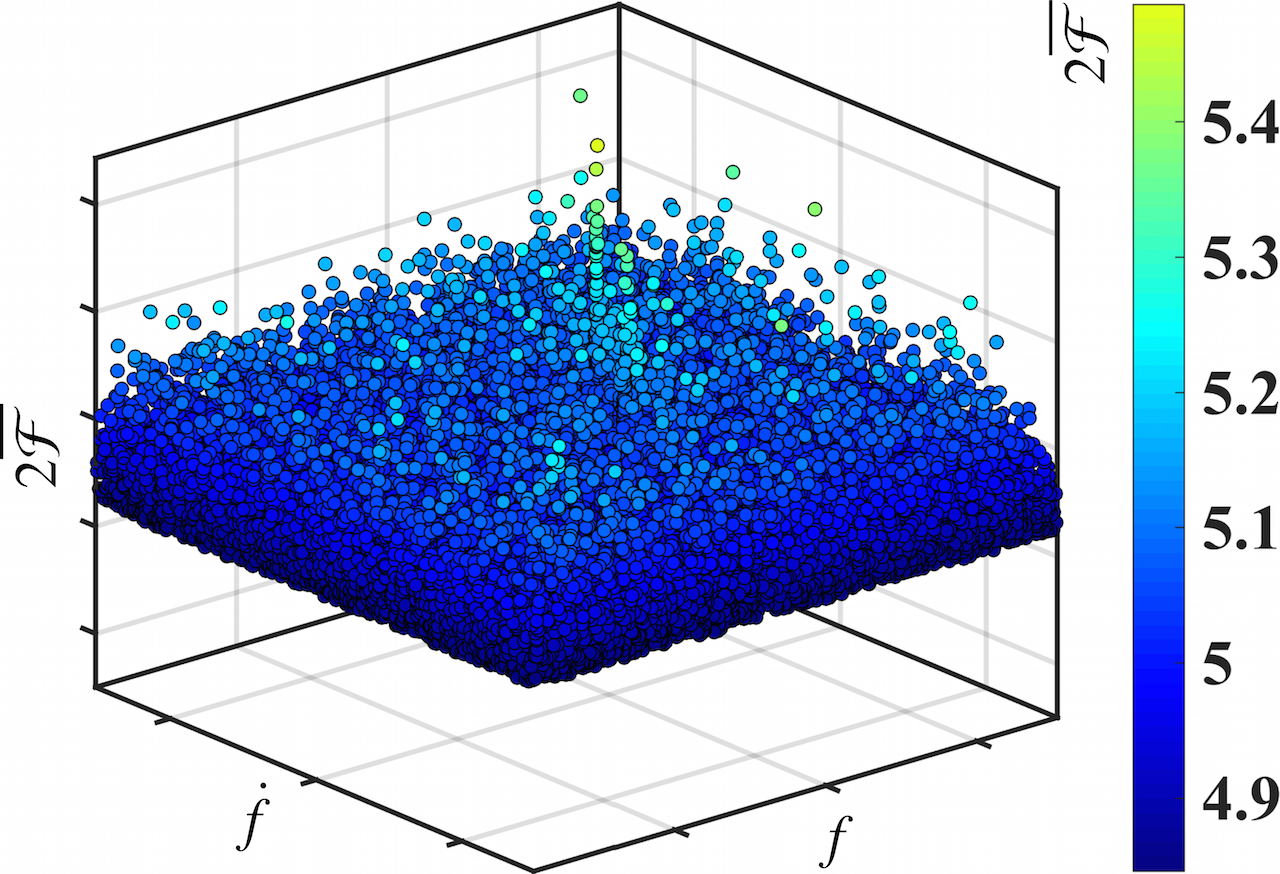}
\caption{{\small We plot the color-coded $2\overline{\mathcal{F}}$ values on the z-axis in three 50$\,$mHz bands. The top-most band is marked as ''disturbed''; the middle band is an example of an ''undisturbed'' band; the bottom-most band is an example of an ''undisturbed'' band but containing a simulated continuous gravitational wave signal.}}
\label{fig:bands}
\end{figure}\vspace{-0.2in}
\begin{figure}[H]
\centering\includegraphics[width=78mm]{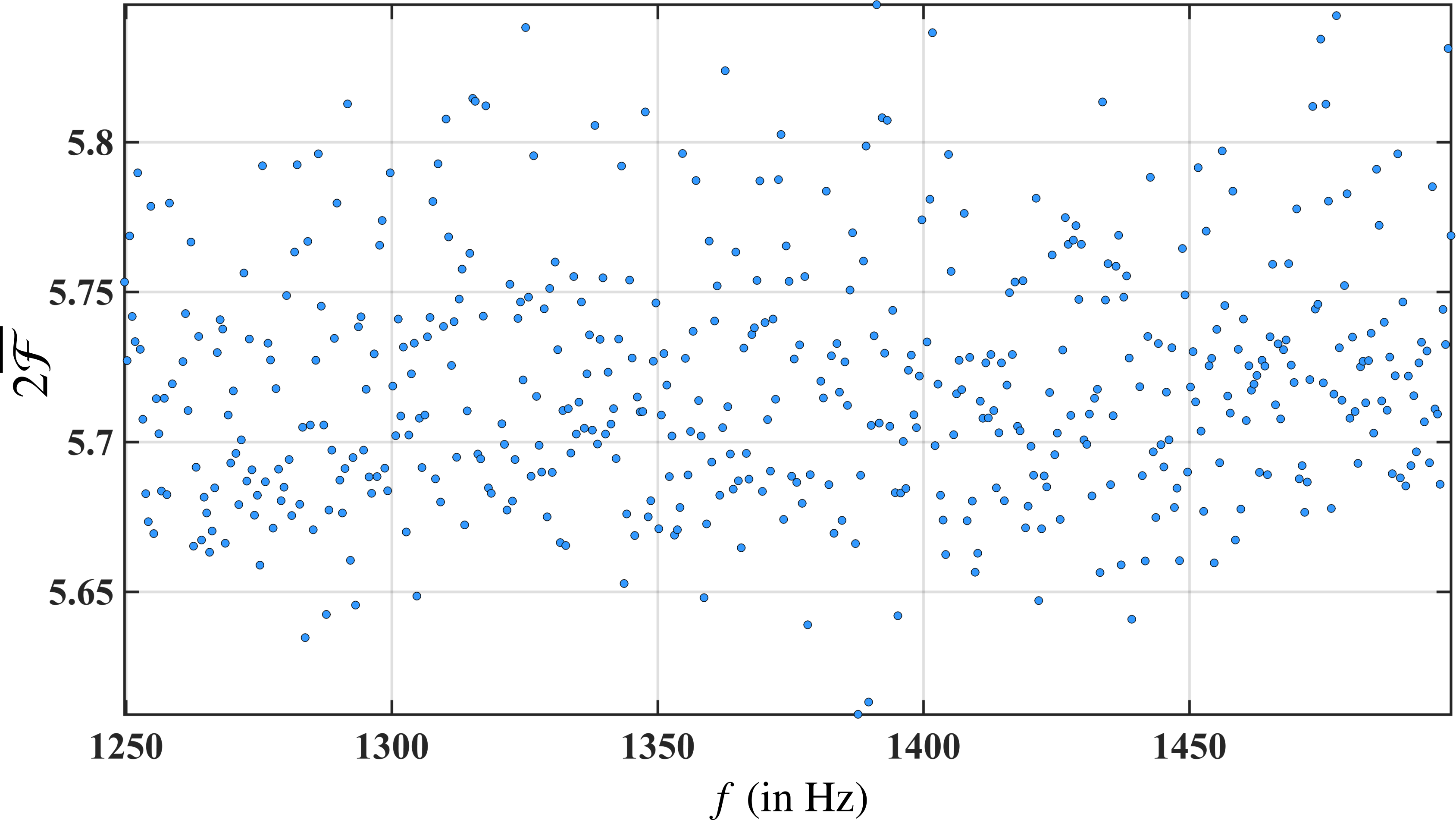}
\caption{{\small Highest values of $2\overline{\mathcal{F}}$ in every 0.5$\,$Hz band as a function of starting frequency of the band.}}
\label{fig:loudest}
\end{figure}
{\noindent}ranking marks the band as ``undisturbed'', a `3' ranks the band as `disturbed'', and rankings of `1' or `2' mark the band as ``marginally disturbed''. A 50$\,$mHz band is eventually considered to be undisturbed if it is marked as '0' by both scientists. The criteria used for this inspection are based on training-sets of real data containing simulated signals. These criteria are designed to exclude disturbed set of results while retaining data sets with signal-like properties, and to err on the side of being conservative in terms of not falsely dismissing signals.A significant part of this visual inspection work can be automated \cite{S6CasA}, but at the time of this search, the procedure had not been fully tested and tuned. In figure \ref{fig:bands}, we empirically illustrate these criteria using three examples. Following this procedure, 3\% of the total 5000 50$\,$mHz bands are marked as ``disturbed" by visual inspection. These excluded bands are listed in table \ref{table:excluded} (Type D), together with the 50$\,$mHz bands excluded as a result of the cleaning of known disturbances above (Type C), i.e. marked as ``All Fake Data'' in table \ref{table:ranges}. In consequence to these exclusions, there exist 0.5$\,$Hz bands comprising results from less than ten 50$\,$mHz bands. We define `fill-level' as the percentage of 50$\,$mHz bands that contribute to the results in 0.5$\,$Hz intervals, where 100\% fill-level signifies contribution by all ten 50$\,$mHz bands. In figure \ref{fig:fill}, we show the distribution of fill-levels for the 0.5$\,$Hz bands searched.

In figure \ref{fig:loudest}, we plot the loudest observed candidate i.e. the candidate with the highest $2\overline{\mathcal{F}}$ value in each 0.5$\,$Hz band in the search frequency range. The loudest candidate in our search has a detection-statistic value of $2\overline{\mathcal{F}}=5.846$ at a frequency of roughly 1391.667$\,$Hz. In order to determine the significance of this loudest candidate, we compare it to the expected value for the highest detection-statistic in our search. In order to determine this expected value, we have to estimate the number of independent trials performed in the search i.e. total number of independent realizations of our detection-statistic $2\overline{\mathcal{F}}$. 

\begin{figure}[H]
\centering\includegraphics[width=78mm]{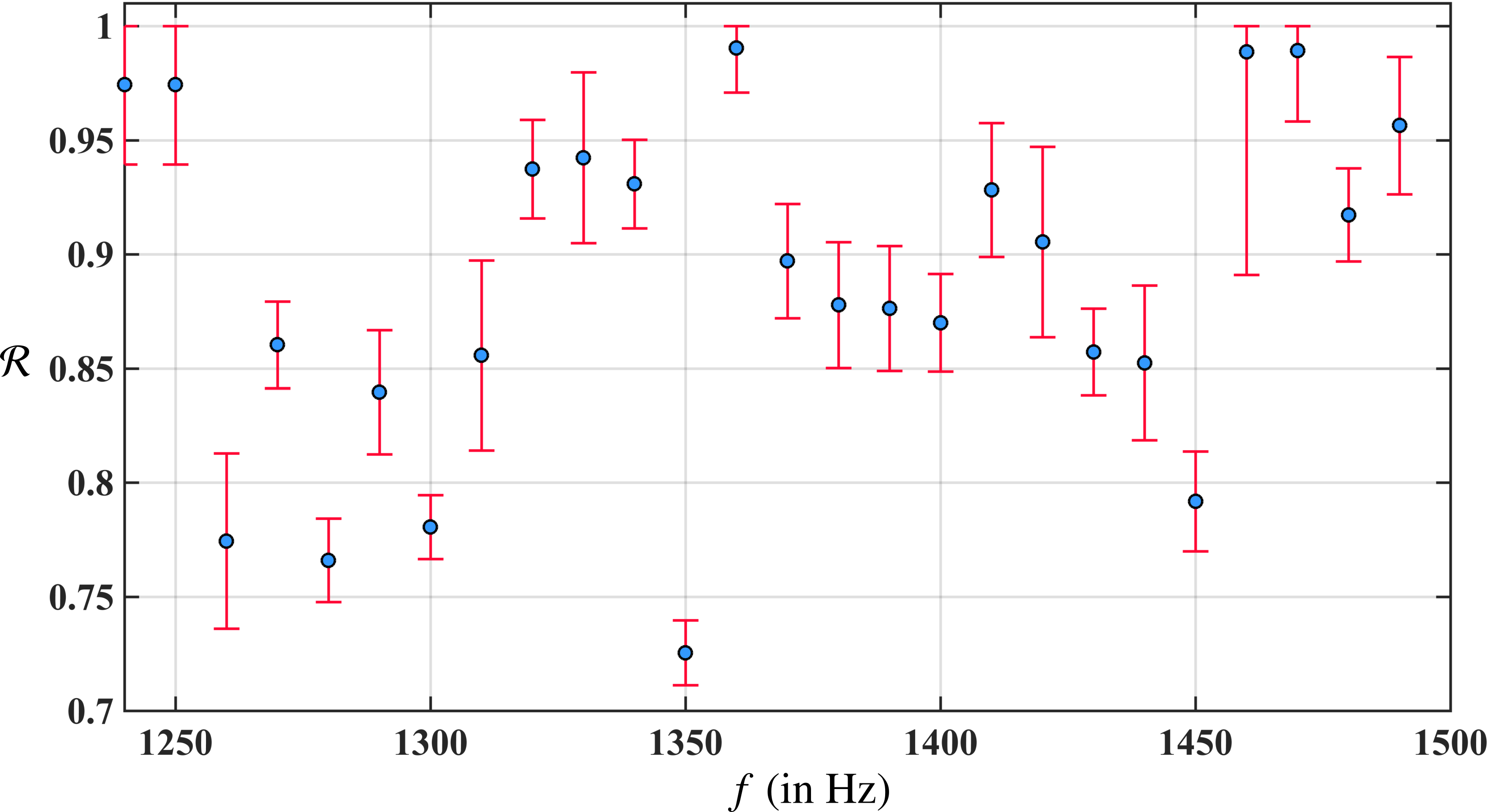}
\caption{\small Plotted ratio $\mathcal{R}=\mathrm{N}_\mathsf{trials}/\mathrm{N}_\mathsf{templates}$ as a function of frequency in 10$\,$Hz intervals. The error bars represent the 1-$\sigma$ statistical errors from the fitting procedure described in the text.} 
\label{fig:ratio}
\end{figure}

\hypertarget{mylink}{\textcolor{black}{The number of independent realizations of the detection-statistic, $\mathrm{N}_\mathsf{trials}$, in a search through a bank of signal templates is smaller than the total number of searched templates, $\mathrm{N}_\mathsf{templates}$. We estimate $\mathrm{N}_\mathsf{trials}$ as a function of frequency in 10$\,$Hz frequency intervals. In each of these 10$\,$Hz intervals, we fit the distribution of loudest candidates from 50$\,$mHz bands to the expected distribution \citep{GalacticCenter}, and obtain the best-fitted value of $\mathrm{N}_\mathsf{trials}$. We perform this calculation in 10$\,$Hz intervals since the sky-grids, along with $\mathrm{N}_\mathsf{templates}$, are constant over 10$\,$Hz frequency intervals. In figure \ref{fig:ratio}, we plot the ratio $\mathcal{R}=\mathrm{N}_\mathsf{trials}/\mathrm{N}_\mathsf{templates}$, as a function of frequency.}}

With $\mathcal{R}(f)$ in hand, we evaluate the expected value for the loudest detection-statistic ($2\overline{\mathcal{F}}_\mathsf{exp}$) in 0.5$\,$Hz bands, and the standard deviation ($\sigma_\mathsf{exp}$) of the associated distribution using (5)-(6) of \citep{GalacticCenter}, with $\mathrm{N}_\mathsf{seg}=205$ and $\mathrm{N}_\mathsf{trials}=\mathcal{R}\,\mathrm{N}_\mathsf{templates}$. Based on these values, we can estimate the significance of the observed loudest candidates (denoted by $2\overline{\mathcal{F}}_{\text{Loud}}$) as the `Critical Ratio' (CR),
\begin{equation}
{\text{CR}}:=\frac{2\overline{\mathcal{F}}_{\mathsf{Loud}}- 2\overline{\mathcal{F}}_{\mathsf{exp}}}{\sigma_{\mathsf{exp}}}.
\label{eq:CR}
\end{equation}\vspace{-0.2in}
\vspace{-0.2in}
\begin{figure}[H]
\centering\includegraphics[width=78mm]{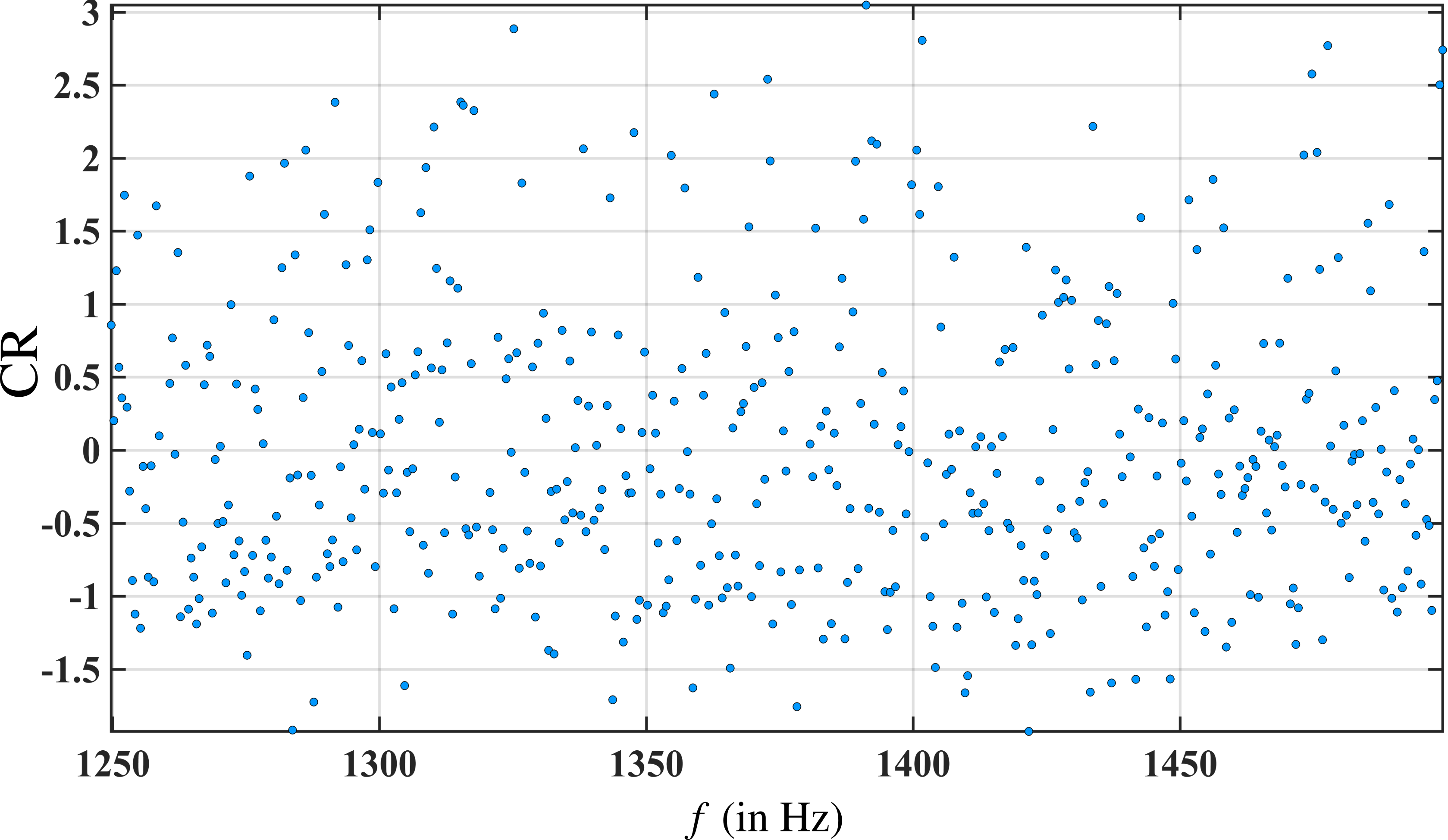}\vspace{0.02in}
\centering\includegraphics[width=78mm]{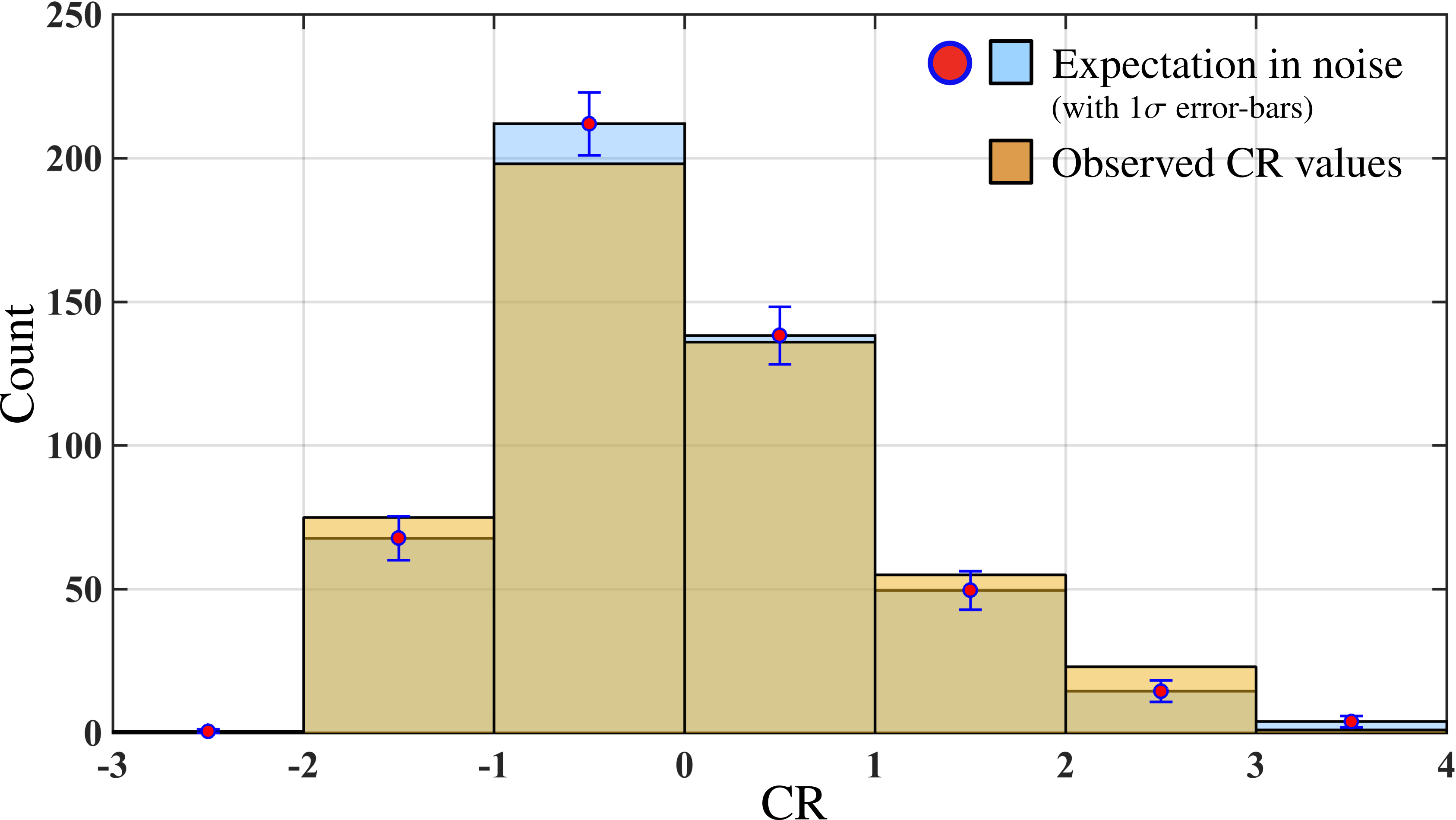}
\caption{{\small In the top panel, we plot the significance of the loudest observed candidate in every 0.5$\,$Hz band as a function of starting frequency of the band. In the bottom panel, we show the distribution of $\text{CR}$ values (top brown histogram bars), and the expected distribution of CR values for pure noise for reference (bottom blue histogram bars with markers). The significance folds in the expected value for the loudest $2\overline{\mathcal{F}}$ and its standard deviation.}}
\label{fig:sigma}
\end{figure}\vspace{-0.1in}
In figure \ref{fig:sigma}, we plot the CR values of the observed loudest candidates in 0.5$\,$Hz bands as a function of frequency (top panel) and their distribution (bottom panel).

\begin{figure}[H]
\centering\includegraphics[width=78mm]{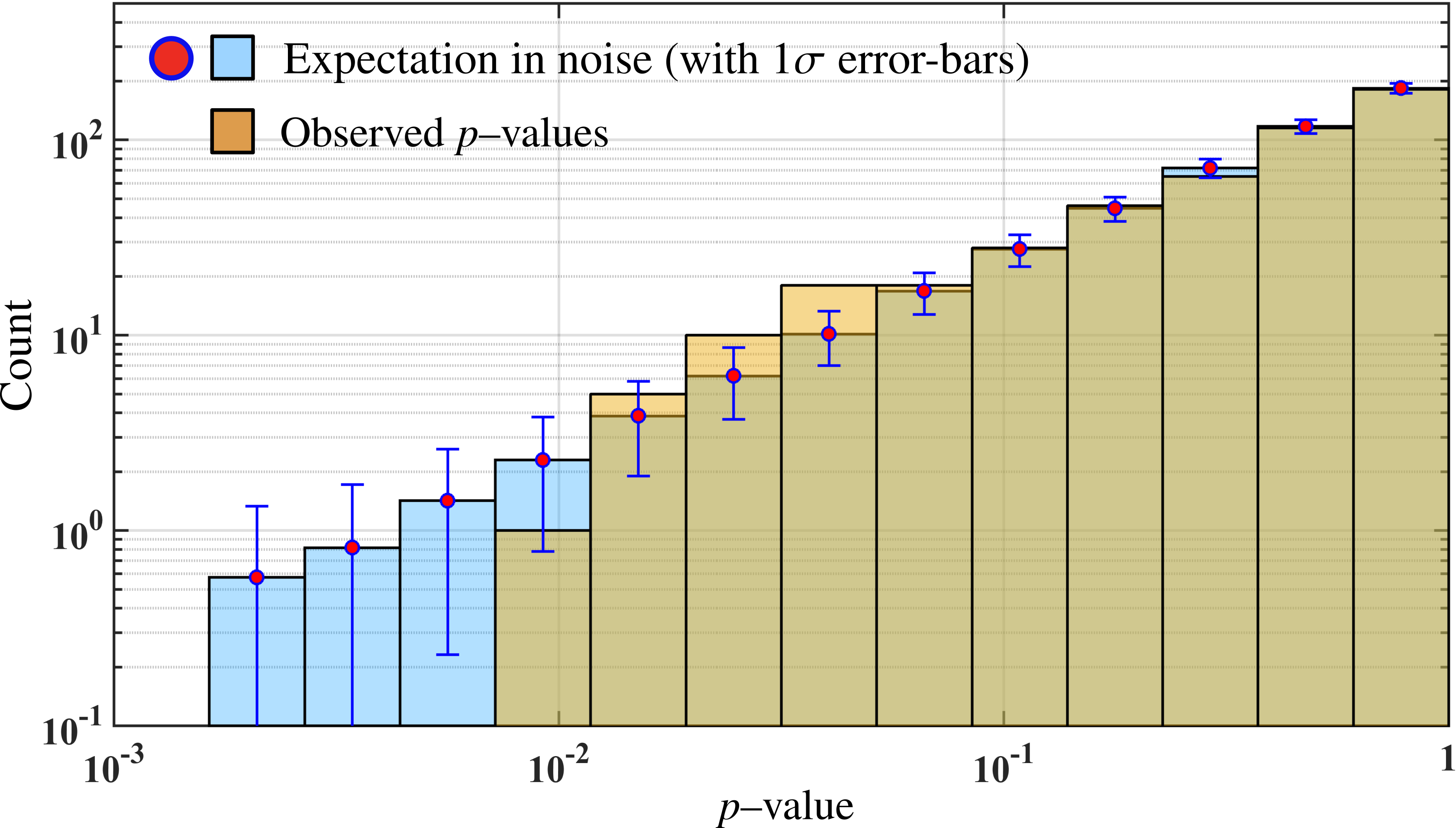}
\caption{{\small $p$-values for the loudest observed candidates in 0.5$\,$Hz bands in the data (top brown histogram bars), and the expected distribution of $p$-values for pure noise for reference (bottom blue histogram bars with markers).}}
\label{fig:pval}
\end{figure}

In this search, the overall loudest candidate with $2\overline{\mathcal{F}}= 5.846$ is also the most significant candidate, with $\text{CR}=3.05$.  A deviation of $3.05\sigma$ from the expected $2\overline{\mathcal{F}}$ value would not be significant enough to claim a detection if we had only searched a single 0.5$\,$Hz band; in fact, it is even less significant considering the fact that a total of 485 0.5$\,$Hz bands were searched.
We define the $p$-value associated with a CR as the probability of observing that particular value of CR or higher by random chance in a search over one 0.5$\,$Hz band, performed over $\mathrm{N}_\mathsf{trials}$ independent trials using $\mathrm{N}_\mathsf{seg}$ segments. In figure \ref{fig:pval}, we see that the distribution of $p$-values associated with the loudest observed candidates in 0.5$\,$Hz bands is consistent with what we expect from the noise-only scenario across the explored parameter space. In particular, we see in figure \ref{fig:pval} that across 485 0.5$\,$Hz bands searched by our set up, we expect $2.3\pm1.5$ candidates at least as significant as $\text{CR}=3.05$ ($p$-value bin $10^{-2}$ for that band) by random chance, which makes our observed loudest candidate completely consistent with expectations from the noise-only case. 
\section{Upper-limits}
\label{sec:ul} 
Our search results do not deviate from the expectations from noise-only data. Hence, we set frequentist upper-limits on the maximum gravitational wave amplitude, $h_\mathrm{0}^{90\%}$, from the target source population consistent with this null result at 90\%-confidence in 0.5$\,$Hz bands. Here, $h_\mathrm{0}^{90\%}$ is the gravitational wave amplitude for which 90\% of the target population of signals would have produced a value of the detection statistic higher than the observed value. 

Hence, we set frequentist upper-limits on the maximum detectable gravitational wave amplitude,  $h_\mathrm{0}^{90\%}$, at 90\%-confidence in 0.5$\,$Hz bands. Here, $h_\mathrm{0}^{90\%}$ is the gravitational wave amplitude for which 90\% of a population of signals with parameter values in our search range would have been produced a value of the detection statistic higher than the observed one in that search range.  

Ideally, in order to estimate the $h_\mathrm{0}^{90\%}$ values in each 0.5$\,$Hz band across the 250$\,$Hz signal-frequency search range, we would perform Monte-Carlo injection-and-recovery simulations in each of those bands. However, this is computationally very intensive. Therefore, we perform Monte-Carlo simulations in six 0.5$\,$Hz bands spread evenly across the 250$\,$Hz-wide frequency range, and in each of these six bands labeled by the index $k$, we estimate the $h_{\mathrm{0},\text{CR}_{\scriptstyle i}}^{90\%,k}$ upper-limit value corresponding to eight different ${\text{CR}}_i$ `significance bins' for the putative observed loudest candidate: $(0.0, 0.5, 1.0, 1.5, 2.0, 2.5, 3.0, 3.5)$.
In each of these six bands and for each of the eight detection criteria, we calculate the so-called `sensitivity-depth', defined in \citep{GalacticCenter}: $\mathcal{D}_{\text{CR}_{\scriptstyle i}}^{90\%,k}$. Lastly, we average these sensitivity-depths over the six bands and derive the average sensitivity-depth $\mathcal{D}_{\text{CR}_{\scriptstyle i}}^{90\%}$ for each detection criterion. The values of the sensitivity-depths range between $\mathcal{D}_{\text{CR}_{\scriptstyle 0.0}}^{90\%}=30.6\,\mathrm{Hz}^{-1/2}$ and $\mathcal{D}_{\text{CR}_{\scriptstyle 3.5}}^{90\%}=28.8\,\mathrm{Hz}^{-1/2}$. We use these $\mathcal{D}_{\text{CR}_i}^{90\%}$ values to set upper-limits in the bands (labeled by $l$) where we have not performed any Monte-Carlo simulations as follows:
\begin{equation}
h_{\mathrm{0}}^{90\%}(f_{l}) = \frac{\sqrt{\mathsf{S_h}(f_{l})}}{\mathcal{D}_{\text{CR}_{\scriptstyle i(l)}}^{90\%}},
\label{eq:depth}
\end{equation}
where, $\text{CR}_i(l)$ is the `significance bin' $i$ corresponding to the loudest observed candidate in the $l$-th frequency band, and $\mathsf{S_h}(f_l)$ is the average amplitude spectral density of the data in that band, measured in Hz$^{-1/2}$. The uncertainties on the $h_\mathrm{0}^{90\%}$ upper-limit values introduced by this procedure amount to roughly 10\% of the nominal $h_\mathrm{0}^{90\%}$ upper-limit value. The final $h_\mathrm{0}^{90\%}$ upper-limit values for this search, including an additional 10\% calibration uncertainty, are given in table \ref{table:ul}, and shown in figure \ref{fig:ul}.  

Note that we do not set upper limits in 0.5$\,$Hz bands where the results are entirely produced with fake Gaussian data inserted by the cleaning procedure described in section \ref{sec:undisturbedbands}; $h_\mathrm{0}^{90\%}$ upper-limit values for such bands do not appear either in table \ref{table:ul}, or in figure \ref{fig:ul}.

Moreover, there also exist 50$\,$mHz bands that contain results contributed by entirely fake data as a result of the cleaning procedure, or that have been excluded from the analysis because they are marked as `disturbed' by the  {{visual inspection method}} described in section \ref{sec:undisturbedbands}. We mark the 0.5$\,$Hz bands which host these particular 50$\,$mHz bands with empty circles in figure \ref{fig:ul}. In table \ref{table:excluded}, we provide a complete list of such 50$\,$mHz bands, highlighting that the upper-limit values do not apply to these bands. Finally, we note that, because of the cleaning procedure, there exist signal-frequency bands where the search results {\fontfamily{ppl}\selectfont \textit{may}} have contributions from some fake data. We list these signal-frequency ranges in table \ref{table:ranges}. In line with the remarks in section \ref{sec:undisturbedbands}, and for the sake of completeness, table \ref{table:ranges} also contains the cleaned bands that featured under Type C in table \ref{table:excluded}, under the column header ``All Fake Data''.

\section{Conclusions}
\label{sec:conclusions}
\begin{figure}[H]
\centering\includegraphics[width=158mm]{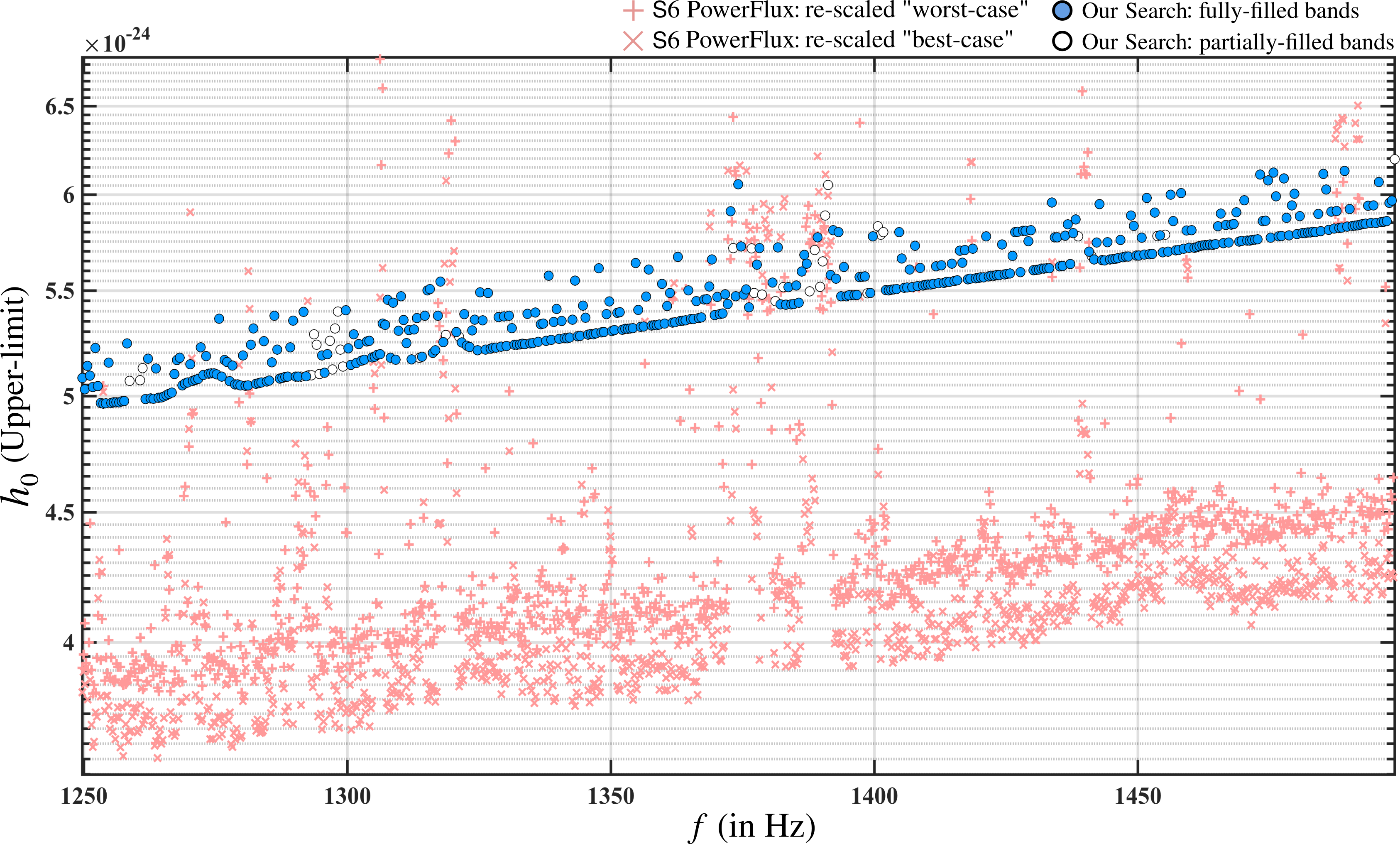}
\caption{{\small 90\%-confidence upper-limits on the gravitational wave amplitude for signals with frequency within 0.5$\,$Hz bands, over the entire sky, and within the spin-down range of the search described in section \ref{sec:Search}. The empty circular markers denote 0.5$\,$Hz bands where the upper-limit value does not hold for all frequencies in that interval; the list of corresponding excluded frequencies is given in table \ref{table:ranges}. 
For reference, we also plot the upper-limit results (with non-circular markers) from the only other high-frequency search \citep{S6PowerFlux}, on significantly more sensitive $\mathsf{S6}$ data. It should be noted that the upper-limits from the {\fontfamily{ppl}\selectfont\textit{PowerFlux}} search \citep{S6PowerFlux} are set at 95\%-confidence rather than 90\%-confidence level as in this search, but refer to 0.25$\,$Hz bands rather than 0.5$\,$Hz bands.}}
\label{fig:ul}
\end{figure}

This search did not yield any evidence of continuous gravitational waves in the LIGO 5th Science Run data in the high-frequency range of 1250--1500$\,$Hz. The lowest value for the upper-limit is  $5.0\times10^{-24}$ for signal frequencies between 1253.217--1255.217$\,$Hz. We show in figure \ref{fig:ul} that these $h_\mathrm{0}^{90\%}$ upper-limits are about 33\% higher than the upper-limits\footnote{The upper-limit values of \citep{S6PowerFlux} have been re-scaled according to \citep{Karl} in order to allow a direct comparison with our $h_\mathrm{0}^{90\%}$ upper-limit results.} set \citep{S6PowerFlux} in the same frequency range but using $\mathsf{S6}$ data .
In this frequency range, the $\mathsf{S6}$ run data is about a factor 2.4 more sensitive compared to the $\mathsf{S5}$ data used in this search. 

\begin{figure}[H]
\centering
\includegraphics[width=78mm]{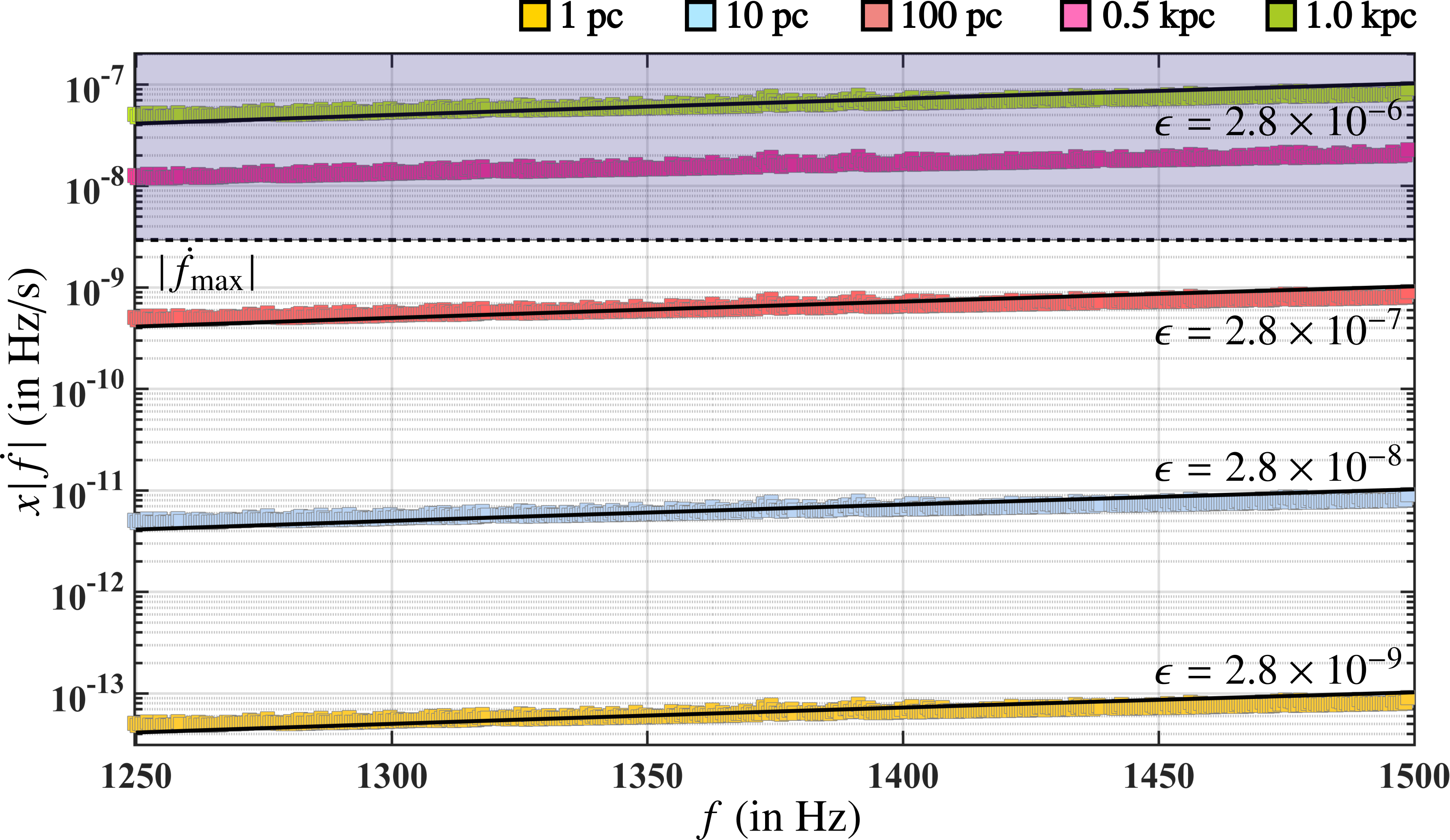}
\includegraphics[width=78mm]{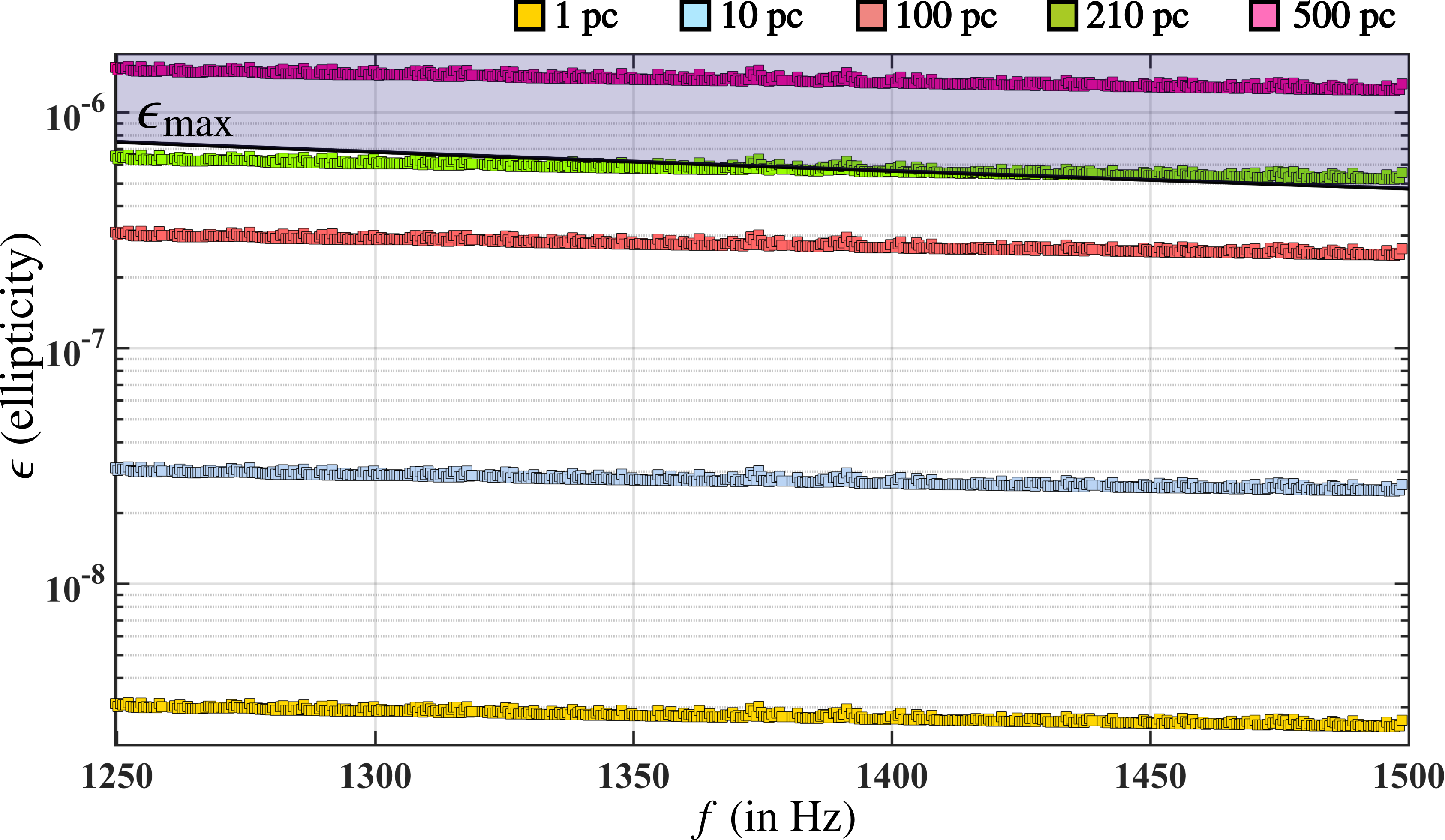}
\caption{{\small Gravitational wave amplitude upper-limits recast as curves in the $\{f,x|\dot{f}|\}$-plane (left panel) for sources at given distances, where $f$ is the signal-frequency and $x|\dot{f}|$ is the gravitational wave spin-down i.e. the fraction of the actual spin-down $|\dot{f}|$ that accounts for the rotational energy loss due to gravitational wave emission. We have superimposed curves of constant ellipticity $\epsilon$. The dotted line at $|\dot{f}_{\textrm{max}}|$ indicates the maximum magnitude of searched spin-down, namely $2.93\times 10^{-9}$ Hz/s. The right panel shows the corresponding $\{f,\epsilon\}$ upper-limit curves for sources at various distances. The $\epsilon_{\textrm{max}}=41.3\times f^{-5/2}$ curve is the ellipticity corresponding to the highest $|\dot{f}|$ searched.}}
\label{fig:astro}
\end{figure}

We can express the $h_\mathrm{0}^{90\%}$ upper-limits as bounds on the maximum distance from Earth within which we can exclude a rotating compact object emitting continuous gravitational waves at a given frequency $f$ due to a fixed and non-axisymmetric mass quadrupole moment, characterised by $\epsilon\mathcal{I}$, with $\mathcal{I}$ being the principal moment of inertia, and $\epsilon$ the ellipticity of the object. 
The `GW-spindown' is the fraction of spin-down, $x|\dot{f}|$, responsible for continuous gravitational wave emission \citep{Jing}. The ellipticity ($\epsilon$) of the compact object necessary to sustain such emission is given by
\begin{equation}
\label{eq:spindown}
\epsilon(f,x|\dot{f}|)=\sqrt{\frac{5c^5}{32\uppi^4\mathrm{G}}\frac{x|\dot{f}|}{\mathcal{I}f^5}},
\end{equation}
{\noindent}where, $c$ is the speed of light, $\mathrm{G}$ is the Gravitational constant. Moreover, since the gravitational wave amplitude for an object at a distance $d_\mathsf{source}$, with an ellipticity $\epsilon$ given by \eqref{eq:spindown}, is expressed as 
\begin{equation}
\label{eq:amp}
h_\mathrm{0}(f, x|\dot{f}|, d_\mathsf{source})=\frac{1}{d_\mathsf{source}}\sqrt{\frac{5\mathcal{I}\mathrm{G}}{2c^3}\frac{x|\dot{f}|}{f}}.
\end{equation}
We can recast the $h_\mathrm{0}^{90\%}$ upper-limit curves as $(f,x|\dot{f}|)$ curves, or as $(f,\epsilon)$ curves, both parametrised by different values of the distance $d_\mathsf{source}$, as shown in figure \ref{fig:astro}. We find that within 100$\,$pc of Earth, our upper-limits exclude objects with ellipticities higher than roughly $\displaystyle{2.8\times10^{-7}\Bigg[\frac{10^{38}\mathrm{kg\,m^2}}{\mathcal{I}}\Bigg]}$, corresponding to GW-spindown values between roughly $4.0\times10^{-10}$ and 
\end{multicols}
\vspace{-15pt}
\begin{multicols}{2}
{\noindent}$1.0\times10^{-9}$ Hz/s. This value is well below the maximum elastic deformation that a relativistic star could sustain, see \citep{JohnsonMcDaniel} and references therein. 

The search presented here is probably the last all-sky search on $\mathsf{S5}$ data, and by inspecting the higher frequency range for continuous gravitational wave emission, it concludes the Einstein@Home observing campaign on this data. Consistent with the recent results on $\mathsf{S6}$ data \citep{S6PowerFlux}, we also find no continuous GW signal in the $\mathsf{S5}$ data. However, mechanisms for transient or intermittent GW emission have been proposed \citep{SinghEkman, PrixGiampanis, DavidTransients} which would not {\fontfamily{ppl}\selectfont \textit{a priori}} exclude a signal that is ``ON'' during the $\mathsf{S5}$ run and ``OFF'' during the $\mathsf{S6}$ run. The estimates for the time-scales, frequencies, and spin-downs of continuous gravitational wave signals from isolated neutron stars lasting weeks to months span a very broad range of values -- orders of magnitude. There are several different mechanisms that could sustain such emission at a level that this search could have detected, and with spin-down values consistent with the total energy emitted in the process, and with the spin-down range spanned by this search.\vspace{-5pt}
\section{Acknowledgments}
\label{sec:ack}
Maria Alessandra Papa, Bruce Allen and Xavier Siemens gratefully acknowledges the support from \textsf{NSF} \textsf{PHY} Grant {\small\qag{1104902}}. All the post-processing computational work for this search was carried out on the $\mathsf{ATLAS}$ super-computing cluster at the Max-Planck-Institut f{\"u}r Gravitationsphysik/ Leibniz Universit{\"a}t Hannover. We also acknowledge the Continuous Wave Group of the LIGO Scientific Collaboration for useful discussions, and in particular, its chair Keith Riles for his careful reading of the manuscript. This paper has been assigned the LIGO Document number {\small\qag{P1600196}}.\begin{center}$\ast\,\ast\,\ast$\end{center}
\end{multicols}
\vspace{-15pt}
\section*{Appendix: Tabular data}
\vspace{-10pt}
\begin{multicols}{2}
\subsection{Upper-limit $\boldsymbol{h_0^{90\%}}$ values}
\label{section:ultable}
\end{multicols}
\vspace{-22pt}
\begin{table}[H]
\begin{center}
\bgroup
\def\arraystretch{0.95}
\begin{tabular}{|@{\hskip 3pt}c@{\hskip 3pt}|@{\hskip 2pt}c@{\hskip 2pt}|c|@{\hskip 3pt}c@{\hskip 3pt}|@{\hskip 3pt}c@{\hskip 3pt}|c|@{\hskip 3pt}c@{\hskip 3pt}|@{\hskip 2pt}c@{\hskip 2pt}|c|@{\hskip 3pt}c@{\hskip 3pt}|@{\hskip 2pt}c@{\hskip 2pt}|}
\hline
\hline
\cline{1-11}\rule{0in}{0.15in}
$\boldsymbol{f}$ \textbf{(in Hz)} & $\boldsymbol{h_0^{90\%}\times 10^{24}}$ & \cellcolor{black!25} & $\boldsymbol{f}$ \textbf{(in Hz)} & $\boldsymbol{h_0^{90\%}\times 10^{24}}$ & \cellcolor{black!25} & $\boldsymbol{f}$ \textbf{(in Hz)} & $\boldsymbol{h_0^{90\%}\times 10^{24}}$ & \cellcolor{black!25} & $\boldsymbol{f}$ \textbf{(in Hz)} & $\boldsymbol{h_0^{90\%}\times 10^{24}}$ \vspace{0.01in}\\
\cline{1-11}\hline\hline\rule{0in}{0.15in}\hspace{-3pt}
1249.717 & 5.1 $\pm$ 1.0 & & 1250.217 & 5.0 $\pm$ 1.0 & & 1250.717 & 5.1 $\pm$ 1.0 & & 1251.217 & 5.1 $\pm$ 1.0 \\
1251.717 & 5.0 $\pm$ 1.0 & & 1252.217 & 5.2 $\pm$ 1.1 & & 1252.717 & 5.0 $\pm$ 1.0 & & 1253.217 & 5.0 $\pm$ 0.9 \\
1253.717 & 5.0 $\pm$ 0.9 & & 1254.217 & 5.0 $\pm$ 0.9 & & 1254.717 & 5.2 $\pm$ 1.0 & & 1255.217 & 5.0 $\pm$ 0.9 \\
1255.717 & 5.0 $\pm$ 0.9 & & 1256.217 & 5.0 $\pm$ 0.9 & & 1256.717 & 5.0 $\pm$ 0.9 & & 1257.217 & 5.0 $\pm$ 0.9 \\
1257.717 & 5.0 $\pm$ 0.9 & & 1258.217 & 5.2 $\pm$ 1.1 & & 1258.717 & 5.1 $\pm$ 1.0 & & 1260.717 & 5.1 $\pm$ 1.0 \\
1261.217 & 5.1 $\pm$ 1.0 & & 1261.717 & 5.0 $\pm$ 0.9 & & 1262.217 & 5.2 $\pm$ 1.0 & & 1262.717 & 5.0 $\pm$ 0.9 \\
1263.217 & 5.0 $\pm$ 0.9 & & 1263.717 & 5.1 $\pm$ 1.0 & & 1264.217 & 5.0 $\pm$ 0.9 & & 1264.717 & 5.0 $\pm$ 0.9 \\
1265.217 & 5.0 $\pm$ 0.9 & & 1265.717 & 5.0 $\pm$ 0.9 & & 1266.217 & 5.0 $\pm$ 0.9 & & 1266.717 & 5.0 $\pm$ 0.9 \\
1267.217 & 5.1 $\pm$ 1.0 & & 1267.717 & 5.2 $\pm$ 1.0 & & 1268.217 & 5.2 $\pm$ 1.0 & & 1268.717 & 5.0 $\pm$ 0.9 \\
1269.217 & 5.1 $\pm$ 0.9 & & 1269.717 & 5.1 $\pm$ 0.9 & & 1270.217 & 5.1 $\pm$ 1.0 & & 1270.717 & 5.1 $\pm$ 0.9 \\
1271.217 & 5.1 $\pm$ 0.9 & & 1271.717 & 5.1 $\pm$ 0.9 & & 1272.217 & 5.2 $\pm$ 1.0 & & 1272.717 & 5.1 $\pm$ 0.9 \\
1273.217 & 5.2 $\pm$ 1.0 & & 1273.717 & 5.1 $\pm$ 0.9 & & 1274.217 & 5.1 $\pm$ 0.9 & & 1274.717 & 5.1 $\pm$ 0.9 \\
1275.217 & 5.1 $\pm$ 0.9 & & 1275.717 & 5.4 $\pm$ 1.1 & & 1276.217 & 5.1 $\pm$ 0.9 & & 1276.717 & 5.2 $\pm$ 1.0 \\
1277.217 & 5.2 $\pm$ 1.0 & & 1277.717 & 5.1 $\pm$ 0.9 & & 1278.217 & 5.1 $\pm$ 1.0 & & 1278.717 & 5.1 $\pm$ 0.9 \\
1279.217 & 5.1 $\pm$ 0.9 & & 1279.717 & 5.0 $\pm$ 0.9 & & 1280.217 & 5.2 $\pm$ 1.0 & & 1280.717 & 5.0 $\pm$ 0.9 \\
1281.217 & 5.0 $\pm$ 0.9 & & 1281.717 & 5.2 $\pm$ 1.0 & & 1282.217 & 5.3 $\pm$ 1.1 & & 1282.717 & 5.1 $\pm$ 0.9 \\
1283.217 & 5.1 $\pm$ 0.9 & & 1283.717 & 5.1 $\pm$ 0.9 & & 1284.217 & 5.3 $\pm$ 1.0 & & 1284.717 & 5.1 $\pm$ 0.9 \\
1285.217 & 5.1 $\pm$ 0.9 & & 1285.717 & 5.2 $\pm$ 1.0 & & 1286.217 & 5.4 $\pm$ 1.1 & & 1286.717 & 5.2 $\pm$ 1.0 \\
1287.217 & 5.1 $\pm$ 0.9 & & 1287.717 & 5.1 $\pm$ 0.9 & & 1288.217 & 5.1 $\pm$ 0.9 & & 1288.717 & 5.1 $\pm$ 0.9 \\
1289.217 & 5.2 $\pm$ 1.0 & & 1289.717 & 5.4 $\pm$ 1.1 & & 1290.217 & 5.1 $\pm$ 0.9 & & 1290.717 & 5.1 $\pm$ 0.9 \\
1291.217 & 5.1 $\pm$ 0.9 & & 1291.717 & 5.4 $\pm$ 1.1 & & 1292.217 & 5.1 $\pm$ 0.9 & & 1292.717 & 5.1 $\pm$ 0.9 \\
1293.217 & 5.1 $\pm$ 0.9 & & 1293.717 & 5.3 $\pm$ 1.0 & & 1294.217 & 5.2 $\pm$ 1.0 & & 1294.717 & 5.1 $\pm$ 0.9 \\
1295.217 & 5.2 $\pm$ 1.0 & & 1295.717 & 5.1 $\pm$ 0.9 & & 1296.217 & 5.2 $\pm$ 1.0 & & 1296.717 & 5.3 $\pm$ 1.0 \\
1297.217 & 5.1 $\pm$ 0.9 & & 1297.717 & 5.3 $\pm$ 1.0 & & 1298.217 & 5.4 $\pm$ 1.1 & & 1298.717 & 5.2 $\pm$ 1.0 \\
1299.217 & 5.1 $\pm$ 0.9 & & 1299.717 & 5.4 $\pm$ 1.1 & & 1300.217 & 5.2 $\pm$ 1.0 & & 1300.717 & 5.1 $\pm$ 0.9 \\
1301.217 & 5.3 $\pm$ 1.0 & & 1301.717 & 5.2 $\pm$ 0.9 & & 1302.217 & 5.2 $\pm$ 1.0 & & 1302.717 & 5.2 $\pm$ 0.9 \\
1303.217 & 5.2 $\pm$ 0.9 & & 1303.717 & 5.3 $\pm$ 1.0 & & 1304.217 & 5.3 $\pm$ 1.0 & & 1304.717 & 5.2 $\pm$ 0.9 \\
1305.217 & 5.2 $\pm$ 0.9 & & 1305.717 & 5.2 $\pm$ 0.9 & & 1306.217 & 5.2 $\pm$ 0.9 & & 1306.717 & 5.3 $\pm$ 1.0 \\
1307.217 & 5.3 $\pm$ 1.0 & & 1307.717 & 5.5 $\pm$ 1.1 & & 1308.217 & 5.2 $\pm$ 0.9 & & 1308.717 & 5.4 $\pm$ 1.1 \\
1309.217 & 5.2 $\pm$ 0.9 & & 1309.717 & 5.3 $\pm$ 1.0 & & 1310.217 & 5.5 $\pm$ 1.1 & & 1310.717 & 5.4 $\pm$ 1.0 \\
1311.217 & 5.2 $\pm$ 1.0 & & 1311.717 & 5.3 $\pm$ 1.0 & & 1312.217 & 5.2 $\pm$ 0.9 & & 1312.717 & 5.3 $\pm$ 1.0 \\
1313.217 & 5.4 $\pm$ 1.0 & & 1313.717 & 5.2 $\pm$ 0.9 & & 1314.217 & 5.2 $\pm$ 0.9 & & 1314.717 & 5.4 $\pm$ 1.0 \\
1315.217 & 5.5 $\pm$ 1.1 & & 1315.717 & 5.5 $\pm$ 1.1 & & 1316.217 & 5.2 $\pm$ 0.9 & & 1316.717 & 5.2 $\pm$ 0.9 \\
1317.217 & 5.4 $\pm$ 1.0 & & 1317.717 & 5.5 $\pm$ 1.1 & & 1318.217 & 5.2 $\pm$ 0.9 & & 1318.717 & 5.3 $\pm$ 0.9 \\
1320.717 & 5.3 $\pm$ 0.9 & & 1321.217 & 5.3 $\pm$ 0.9 & & 1321.717 & 5.2 $\pm$ 0.9 & & 1322.217 & 5.4 $\pm$ 1.0 \\
1322.717 & 5.2 $\pm$ 0.9 & & 1323.217 & 5.2 $\pm$ 0.9 & & 1323.717 & 5.3 $\pm$ 1.0 & & 1324.217 & 5.4 $\pm$ 1.0 \\
1324.717 & 5.2 $\pm$ 0.9 & & 1325.217 & 5.5 $\pm$ 1.1 & & 1325.717 & 5.4 $\pm$ 1.0 & & 1326.217 & 5.2 $\pm$ 0.9 \\
1326.717 & 5.5 $\pm$ 1.1 & & 1327.217 & 5.2 $\pm$ 0.9 & & 1327.717 & 5.2 $\pm$ 0.9 & & 1328.217 & 5.2 $\pm$ 0.9 \\
1328.717 & 5.4 $\pm$ 1.0 & & 1329.217 & 5.2 $\pm$ 0.9 & & 1329.717 & 5.4 $\pm$ 1.0 & & 1330.217 & 5.2 $\pm$ 0.9 \\
\cline{1-11}
\end{tabular}
\egroup
\end{center}
\end{table}
\pagebreak
\begin{table}[H]
\begin{center}
\bgroup
\def\arraystretch{0.95}
\begin{tabular}{|@{\hskip 3pt}c@{\hskip 3pt}|@{\hskip 2pt}c@{\hskip 2pt}|c|@{\hskip 3pt}c@{\hskip 3pt}|@{\hskip 2pt}c@{\hskip 2pt}|c|@{\hskip 3pt}c@{\hskip 3pt}|@{\hskip 2pt}c@{\hskip 2pt}|c|@{\hskip 3pt}c@{\hskip 3pt}|@{\hskip 2pt}c@{\hskip 2pt}|}
\hline
\hline
\cline{1-11}\rule{0in}{0.15in}
$\boldsymbol{f}$ \textbf{(in Hz)} & $\boldsymbol{h_0^{90\%}\times 10^{24}}$ & \cellcolor{black!25} & $\boldsymbol{f}$ \textbf{(in Hz)} & $\boldsymbol{h_0^{90\%}\times 10^{24}}$ & \cellcolor{black!25} & $\boldsymbol{f}$ \textbf{(in Hz)} & $\boldsymbol{h_0^{90\%}\times 10^{24}}$ & \cellcolor{black!25} & $\boldsymbol{f}$ \textbf{(in Hz)} & $\boldsymbol{h_0^{90\%}\times 10^{24}}$ \vspace{0.01in}\\
\cline{1-11}\hline\hline\rule{0in}{0.15in}\hspace{-3pt}
1330.717 & 5.4 $\pm$ 1.0 & & 1331.217 & 5.3 $\pm$ 1.0 & & 1331.717 & 5.2 $\pm$ 0.9 & & 1332.217 & 5.2 $\pm$ 0.9 \\
1332.717 & 5.2 $\pm$ 0.9 & & 1333.217 & 5.2 $\pm$ 0.9 & & 1333.717 & 5.2 $\pm$ 0.9 & & 1334.217 & 5.4 $\pm$ 1.0 \\
1334.717 & 5.2 $\pm$ 0.9 & & 1335.217 & 5.2 $\pm$ 0.9 & & 1335.717 & 5.4 $\pm$ 1.0 & & 1336.217 & 5.3 $\pm$ 0.9 \\
1336.717 & 5.3 $\pm$ 1.0 & & 1337.217 & 5.3 $\pm$ 1.0 & & 1337.717 & 5.3 $\pm$ 0.9 & & 1338.217 & 5.6 $\pm$ 1.1 \\
1338.717 & 5.3 $\pm$ 0.9 & & 1339.217 & 5.3 $\pm$ 1.0 & & 1339.717 & 5.4 $\pm$ 1.0 & & 1340.217 & 5.3 $\pm$ 0.9 \\
1340.717 & 5.4 $\pm$ 1.0 & & 1341.217 & 5.3 $\pm$ 0.9 & & 1341.717 & 5.3 $\pm$ 0.9 & & 1342.217 & 5.3 $\pm$ 0.9 \\
1342.717 & 5.4 $\pm$ 1.0 & & 1343.217 & 5.6 $\pm$ 1.1 & & 1343.717 & 5.3 $\pm$ 0.9 & & 1344.217 & 5.3 $\pm$ 0.9 \\
1344.717 & 5.4 $\pm$ 1.0 & & 1345.217 & 5.4 $\pm$ 1.0 & & 1345.717 & 5.3 $\pm$ 0.9 & & 1346.217 & 5.3 $\pm$ 0.9 \\
1346.717 & 5.3 $\pm$ 0.9 & & 1347.217 & 5.3 $\pm$ 0.9 & & 1347.717 & 5.6 $\pm$ 1.2 & & 1348.217 & 5.3 $\pm$ 0.9 \\
1348.717 & 5.3 $\pm$ 0.9 & & 1349.217 & 5.4 $\pm$ 1.0 & & 1349.717 & 5.4 $\pm$ 1.0 & & 1350.217 & 5.3 $\pm$ 0.9 \\
1350.717 & 5.3 $\pm$ 0.9 & & 1351.217 & 5.4 $\pm$ 1.0 & & 1351.717 & 5.4 $\pm$ 1.0 & & 1352.217 & 5.3 $\pm$ 0.9 \\
1352.717 & 5.3 $\pm$ 0.9 & & 1353.217 & 5.3 $\pm$ 0.9 & & 1353.717 & 5.3 $\pm$ 0.9 & & 1354.217 & 5.3 $\pm$ 0.9 \\
1354.717 & 5.6 $\pm$ 1.2 & & 1355.217 & 5.4 $\pm$ 1.0 & & 1355.717 & 5.3 $\pm$ 0.9 & & 1356.217 & 5.3 $\pm$ 0.9 \\
1356.717 & 5.5 $\pm$ 1.1 & & 1357.217 & 5.6 $\pm$ 1.1 & & 1357.717 & 5.3 $\pm$ 0.9 & & 1358.217 & 5.3 $\pm$ 0.9 \\
1358.717 & 5.3 $\pm$ 0.9 & & 1359.217 & 5.3 $\pm$ 0.9 & & 1359.717 & 5.5 $\pm$ 1.1 & & 1360.217 & 5.3 $\pm$ 0.9 \\
1360.717 & 5.4 $\pm$ 1.0 & & 1361.217 & 5.5 $\pm$ 1.1 & & 1361.717 & 5.3 $\pm$ 0.9 & & 1362.217 & 5.3 $\pm$ 0.9 \\
1362.717 & 5.7 $\pm$ 1.2 & & 1363.217 & 5.4 $\pm$ 0.9 & & 1363.717 & 5.4 $\pm$ 0.9 & & 1364.217 & 5.4 $\pm$ 0.9 \\
1364.717 & 5.5 $\pm$ 1.1 & & 1365.217 & 5.4 $\pm$ 0.9 & & 1365.717 & 5.4 $\pm$ 0.9 & & 1366.217 & 5.4 $\pm$ 1.1 \\
1366.717 & 5.4 $\pm$ 0.9 & & 1367.217 & 5.4 $\pm$ 0.9 & & 1367.717 & 5.5 $\pm$ 1.1 & & 1368.217 & 5.5 $\pm$ 1.1 \\
1368.717 & 5.5 $\pm$ 1.1 & & 1369.217 & 5.7 $\pm$ 1.1 & & 1369.717 & 5.4 $\pm$ 0.9 & & 1370.217 & 5.5 $\pm$ 1.1 \\
1370.717 & 5.4 $\pm$ 0.9 & & 1371.217 & 5.4 $\pm$ 0.9 & & 1371.717 & 5.5 $\pm$ 1.1 & & 1372.217 & 5.4 $\pm$ 0.9 \\
1372.717 & 5.9 $\pm$ 1.1 & & 1373.217 & 5.7 $\pm$ 1.2 & & 1373.717 & 5.5 $\pm$ 1.0 & & 1374.217 & 6.1 $\pm$ 1.2 \\
1374.717 & 5.7 $\pm$ 1.1 & & 1375.217 & 5.5 $\pm$ 1.0 & & 1375.717 & 5.5 $\pm$ 1.1 & & 1376.217 & 5.4 $\pm$ 0.9 \\
1376.717 & 5.7 $\pm$ 1.1 & & 1377.217 & 5.5 $\pm$ 1.0 & & 1377.717 & 5.6 $\pm$ 1.1 & & 1378.217 & 5.7 $\pm$ 1.0 \\
1378.717 & 5.5 $\pm$ 1.0 & & 1380.717 & 5.5 $\pm$ 1.1 & & 1381.217 & 5.4 $\pm$ 0.9 & & 1381.717 & 5.7 $\pm$ 1.2 \\
1382.217 & 5.4 $\pm$ 0.9 & & 1382.717 & 5.5 $\pm$ 1.1 & & 1383.217 & 5.4 $\pm$ 0.9 & & 1383.717 & 5.5 $\pm$ 1.1 \\
1384.217 & 5.4 $\pm$ 0.9 & & 1384.717 & 5.4 $\pm$ 0.9 & & 1385.217 & 5.5 $\pm$ 1.1 & & 1385.717 & 5.4 $\pm$ 0.9 \\
1386.217 & 5.6 $\pm$ 1.1 & & 1386.717 & 5.7 $\pm$ 1.1 & & 1387.217 & 5.6 $\pm$ 1.0 & & 1387.717 & 5.5 $\pm$ 1.0 \\
1388.717 & 5.7 $\pm$ 1.1 & & 1389.217 & 5.8 $\pm$ 1.2 & & 1389.717 & 5.5 $\pm$ 1.0 & & 1390.217 & 5.6 $\pm$ 1.1 \\
1390.717 & 5.9 $\pm$ 1.2 & & 1391.217 & 6.1 $\pm$ 1.1 & & 1391.717 & 5.6 $\pm$ 1.0 & & 1392.217 & 5.8 $\pm$ 1.2 \\
1392.717 & 5.6 $\pm$ 1.1 & & 1393.217 & 5.8 $\pm$ 1.2 & & 1393.717 & 5.5 $\pm$ 1.0 & & 1394.217 & 5.6 $\pm$ 1.1 \\
1394.717 & 5.5 $\pm$ 1.0 & & 1395.217 & 5.5 $\pm$ 1.0 & & 1395.717 & 5.5 $\pm$ 1.0 & & 1396.217 & 5.5 $\pm$ 1.0 \\
1396.717 & 5.5 $\pm$ 1.0 & & 1397.217 & 5.6 $\pm$ 1.1 & & 1397.717 & 5.6 $\pm$ 1.1 & & 1398.217 & 5.6 $\pm$ 1.1 \\
1398.717 & 5.5 $\pm$ 1.0 & & 1399.217 & 5.5 $\pm$ 1.0 & & 1399.717 & 5.8 $\pm$ 1.2 & & 1400.717 & 5.8 $\pm$ 1.2 \\
1401.217 & 5.8 $\pm$ 1.2 & & 1401.717 & 5.8 $\pm$ 1.1 & & 1402.217 & 5.5 $\pm$ 1.0 & & 1402.717 & 5.5 $\pm$ 1.0 \\
1403.217 & 5.5 $\pm$ 1.0 & & 1403.717 & 5.5 $\pm$ 1.0 & & 1404.217 & 5.5 $\pm$ 1.0 & & 1404.717 & 5.8 $\pm$ 1.2 \\
1405.217 & 5.7 $\pm$ 1.1 & & 1405.717 & 5.5 $\pm$ 1.0 & & 1406.217 & 5.5 $\pm$ 1.0 & & 1406.717 & 5.6 $\pm$ 1.1 \\
1407.217 & 5.5 $\pm$ 1.0 & & 1407.717 & 5.7 $\pm$ 1.1 & & 1408.217 & 5.5 $\pm$ 1.0 & & 1408.717 & 5.6 $\pm$ 1.1 \\
1409.217 & 5.5 $\pm$ 1.0 & & 1409.717 & 5.5 $\pm$ 1.0 & & 1410.217 & 5.5 $\pm$ 1.0 & & 1410.717 & 5.5 $\pm$ 1.0 \\
1411.217 & 5.5 $\pm$ 1.0 & & 1411.717 & 5.6 $\pm$ 1.1 & & 1412.217 & 5.5 $\pm$ 1.0 & & 1412.717 & 5.6 $\pm$ 1.1 \\
1413.217 & 5.5 $\pm$ 1.0 & & 1413.717 & 5.5 $\pm$ 1.0 & & 1414.217 & 5.5 $\pm$ 1.0 & & 1414.717 & 5.6 $\pm$ 1.1 \\
1415.217 & 5.5 $\pm$ 1.0 & & 1415.717 & 5.6 $\pm$ 1.0 & & 1416.217 & 5.7 $\pm$ 1.1 & & 1416.717 & 5.6 $\pm$ 1.1 \\
1417.217 & 5.7 $\pm$ 1.1 & & 1417.717 & 5.6 $\pm$ 1.0 & & 1418.217 & 5.6 $\pm$ 1.0 & & 1418.717 & 5.7 $\pm$ 1.1 \\
1419.217 & 5.6 $\pm$ 1.0 & & 1419.717 & 5.6 $\pm$ 1.0 & & 1420.217 & 5.6 $\pm$ 1.0 & & 1420.717 & 5.6 $\pm$ 1.0 \\
1421.217 & 5.8 $\pm$ 1.1 & & 1421.717 & 5.6 $\pm$ 1.0 & & 1422.217 & 5.6 $\pm$ 1.0 & & 1422.717 & 5.6 $\pm$ 1.0 \\
1423.217 & 5.6 $\pm$ 1.0 & & 1423.717 & 5.6 $\pm$ 1.0 & & 1424.217 & 5.7 $\pm$ 1.1 & & 1424.717 & 5.6 $\pm$ 1.0 \\
1425.217 & 5.6 $\pm$ 1.0 & & 1425.717 & 5.6 $\pm$ 1.0 & & 1426.217 & 5.7 $\pm$ 1.1 & & 1426.717 & 5.8 $\pm$ 1.1 \\
1427.217 & 5.8 $\pm$ 1.1 & & 1427.717 & 5.6 $\pm$ 1.0 & & 1428.217 & 5.8 $\pm$ 1.1 & & 1428.717 & 5.8 $\pm$ 1.1 \\
1429.217 & 5.8 $\pm$ 1.1 & & 1429.717 & 5.8 $\pm$ 1.1 & & 1430.217 & 5.6 $\pm$ 1.0 & & 1430.717 & 5.6 $\pm$ 1.0 \\
1431.217 & 5.6 $\pm$ 1.0 & & 1431.717 & 5.6 $\pm$ 1.0 & & 1432.217 & 5.6 $\pm$ 1.0 & & 1432.717 & 5.6 $\pm$ 1.0 \\
1433.217 & 5.6 $\pm$ 1.0 & & 1433.717 & 6.0 $\pm$ 1.2 & & 1434.217 & 5.8 $\pm$ 1.1 & & 1434.717 & 5.8 $\pm$ 1.1 \\
1435.217 & 5.6 $\pm$ 1.0 & & 1435.717 & 5.6 $\pm$ 1.0 & & 1436.217 & 5.8 $\pm$ 1.1 & & 1436.717 & 5.8 $\pm$ 1.1 \\
1437.217 & 5.6 $\pm$ 1.0 & & 1437.717 & 5.8 $\pm$ 1.1 & & 1438.217 & 5.9 $\pm$ 1.1 & & 1438.717 & 5.8 $\pm$ 1.1 \\
1440.717 & 5.7 $\pm$ 1.0 & & 1441.217 & 5.7 $\pm$ 1.0 & & 1441.717 & 5.7 $\pm$ 1.0 & & 1442.217 & 5.7 $\pm$ 1.1 \\
1442.717 & 5.9 $\pm$ 1.2 & & 1443.217 & 5.7 $\pm$ 1.0 & & 1443.717 & 5.7 $\pm$ 1.0 & & 1444.217 & 5.7 $\pm$ 1.1 \\
1444.717 & 5.7 $\pm$ 1.0 & & 1445.217 & 5.7 $\pm$ 1.0 & & 1445.717 & 5.7 $\pm$ 1.0 & & 1446.217 & 5.7 $\pm$ 1.0 \\
1446.717 & 5.8 $\pm$ 1.1 & & 1447.217 & 5.7 $\pm$ 1.0 & & 1447.717 & 5.7 $\pm$ 1.0 & & 1448.217 & 5.7 $\pm$ 1.0 \\
1448.717 & 5.9 $\pm$ 1.1 & & 1449.217 & 5.8 $\pm$ 1.1 & & 1449.717 & 5.7 $\pm$ 1.0 & & 1450.217 & 5.7 $\pm$ 1.0 \\
1450.717 & 5.8 $\pm$ 1.1 & & 1451.217 & 5.7 $\pm$ 1.0 & & 1451.717 & 6.0 $\pm$ 1.2 & & 1452.217 & 5.7 $\pm$ 1.0 \\
\cline{1-11}
\end{tabular}
\egroup
\end{center}
\end{table}
\pagebreak
\begin{table}[H]
\begin{center}
\bgroup
\def\arraystretch{0.95}
\begin{tabular}{|@{\hskip 3pt}c@{\hskip 3pt}|@{\hskip 2pt}c@{\hskip 2pt}|c|@{\hskip 3pt}c@{\hskip 3pt}|@{\hskip 2pt}c@{\hskip 2pt}|c|@{\hskip 3pt}c@{\hskip 3pt}|@{\hskip 2pt}c@{\hskip 2pt}|c|@{\hskip 3pt}c@{\hskip 3pt}|@{\hskip 2pt}c@{\hskip 2pt}|}
\hline
\hline
\cline{1-11}\rule{0in}{0.15in}
$\boldsymbol{f}$ \textbf{(in Hz)} & $\boldsymbol{h_0^{90\%}\times 10^{24}}$ & \cellcolor{black!25} & $\boldsymbol{f}$ \textbf{(in Hz)} & $\boldsymbol{h_0^{90\%}\times 10^{24}}$ & \cellcolor{black!25} & $\boldsymbol{f}$ \textbf{(in Hz)} & $\boldsymbol{h_0^{90\%}\times 10^{24}}$ & \cellcolor{black!25} & $\boldsymbol{f}$ \textbf{(in Hz)} & $\boldsymbol{h_0^{90\%}\times 10^{24}}$ \vspace{0.01in}\\
\cline{1-11}\hline\hline\rule{0in}{0.15in}\hspace{-3pt}
1452.717 & 5.7 $\pm$ 1.0 & & 1453.217 & 5.9 $\pm$ 1.1 & & 1453.717 & 5.8 $\pm$ 1.1 & & 1454.217 & 5.8 $\pm$ 1.1 \\
1454.717 & 5.7 $\pm$ 1.0 & & 1455.217 & 5.8 $\pm$ 1.1 & & 1455.717 & 5.7 $\pm$ 1.0 & & 1456.217 & 6.0 $\pm$ 1.2 \\
1456.717 & 5.9 $\pm$ 1.1 & & 1457.217 & 5.7 $\pm$ 1.0 & & 1457.717 & 5.7 $\pm$ 1.0 & & 1458.217 & 6.0 $\pm$ 1.2 \\
1458.717 & 5.7 $\pm$ 1.0 & & 1459.217 & 5.8 $\pm$ 1.1 & & 1459.717 & 5.7 $\pm$ 1.0 & & 1460.217 & 5.8 $\pm$ 1.1 \\
1460.717 & 5.7 $\pm$ 1.0 & & 1461.217 & 5.7 $\pm$ 1.0 & & 1461.717 & 5.7 $\pm$ 1.0 & & 1462.217 & 5.7 $\pm$ 1.0 \\
1462.717 & 5.7 $\pm$ 1.0 & & 1463.217 & 5.7 $\pm$ 1.0 & & 1463.717 & 5.7 $\pm$ 1.0 & & 1464.217 & 5.7 $\pm$ 1.0 \\
1464.717 & 5.7 $\pm$ 1.0 & & 1465.217 & 5.8 $\pm$ 1.1 & & 1465.717 & 5.9 $\pm$ 1.1 & & 1466.217 & 5.7 $\pm$ 1.0 \\
1466.717 & 5.8 $\pm$ 1.1 & & 1467.217 & 5.7 $\pm$ 1.0 & & 1467.717 & 5.8 $\pm$ 1.1 & & 1468.217 & 5.8 $\pm$ 1.1 \\
1468.717 & 5.9 $\pm$ 1.1 & & 1469.217 & 5.8 $\pm$ 1.0 & & 1469.717 & 5.8 $\pm$ 1.0 & & 1470.217 & 6.0 $\pm$ 1.2 \\
1470.717 & 5.8 $\pm$ 1.0 & & 1471.217 & 5.8 $\pm$ 1.0 & & 1471.717 & 5.8 $\pm$ 1.0 & & 1472.217 & 5.8 $\pm$ 1.0 \\
1472.717 & 5.8 $\pm$ 1.0 & & 1473.217 & 6.1 $\pm$ 1.3 & & 1473.717 & 5.9 $\pm$ 1.1 & & 1474.217 & 5.9 $\pm$ 1.1 \\
1474.717 & 6.1 $\pm$ 1.2 & & 1475.217 & 5.8 $\pm$ 1.0 & & 1475.717 & 6.1 $\pm$ 1.3 & & 1476.217 & 6.0 $\pm$ 1.2 \\
1476.717 & 5.8 $\pm$ 1.0 & & 1477.217 & 5.8 $\pm$ 1.0 & & 1477.717 & 6.1 $\pm$ 1.2 & & 1478.217 & 5.9 $\pm$ 1.1 \\
1478.717 & 5.8 $\pm$ 1.0 & & 1479.217 & 5.9 $\pm$ 1.1 & & 1479.717 & 6.0 $\pm$ 1.2 & & 1480.217 & 5.8 $\pm$ 1.0 \\
1480.717 & 5.9 $\pm$ 1.1 & & 1481.217 & 5.8 $\pm$ 1.0 & & 1481.717 & 5.8 $\pm$ 1.0 & & 1482.217 & 5.8 $\pm$ 1.0 \\
1482.717 & 5.8 $\pm$ 1.0 & & 1483.217 & 5.8 $\pm$ 1.0 & & 1483.717 & 5.8 $\pm$ 1.0 & & 1484.217 & 5.9 $\pm$ 1.1 \\
1484.717 & 5.8 $\pm$ 1.0 & & 1485.217 & 6.1 $\pm$ 1.2 & & 1485.717 & 6.0 $\pm$ 1.2 & & 1486.217 & 5.8 $\pm$ 1.0 \\
1486.717 & 5.9 $\pm$ 1.1 & & 1487.217 & 5.8 $\pm$ 1.0 & & 1487.717 & 5.9 $\pm$ 1.1 & & 1488.217 & 5.8 $\pm$ 1.0 \\
1488.717 & 5.8 $\pm$ 1.0 & & 1489.217 & 6.1 $\pm$ 1.2 & & 1489.717 & 5.8 $\pm$ 1.0 & & 1490.217 & 5.9 $\pm$ 1.1 \\
1490.717 & 5.8 $\pm$ 1.0 & & 1491.217 & 5.8 $\pm$ 1.0 & & 1491.717 & 5.8 $\pm$ 1.0 & & 1492.217 & 5.8 $\pm$ 1.0 \\
1492.717 & 5.8 $\pm$ 1.0 & & 1493.217 & 5.8 $\pm$ 1.0 & & 1493.717 & 5.9 $\pm$ 1.1 & & 1494.217 & 5.8 $\pm$ 1.0 \\
1494.717 & 5.9 $\pm$ 1.1 & & 1495.217 & 5.8 $\pm$ 1.0 & & 1495.717 & 6.1 $\pm$ 1.2 & & 1496.217 & 5.9 $\pm$ 1.0 \\
1496.717 & 5.9 $\pm$ 1.0 & & 1497.217 & 5.9 $\pm$ 1.0 & & 1497.717 & 6.0 $\pm$ 1.2 & & 1498.217 & 6.0 $\pm$ 1.2 \\
1498.717 & 6.2 $\pm$ 1.3 & &    --    &       --      & &    --    &       --      & &    --    &       --      \\
\cline{1-11}
\end{tabular}
\egroup
\end{center}
\caption{Left column denotes the starting frequency of each 0.5$\,$Hz signal-frequency band in which we set upper-limits; right column states the upper-limit value i.e. $h_0^{90\%}$, for that 0.5$\,$Hz band. Note: the $h_0^{90\%}$ values quoted here include additional 10\% uncertainty introduced by data calibration procedure.}
\label{table:ul}
\end{table}
\subsection{Detector Lines}
\label{section:lines}
\begin{table}[ht!]
\begin{center}
\bgroup
\def\arraystretch{0.95}
\begin{tabular}{|@{\hskip 0.1in}c@{\hskip 0.1in}|@{\hskip 0.1in}c@{\hskip 0.1in}|@{\hskip 0.1in}c@{\hskip 0.1in}|@{\hskip 0.1in}c@{\hskip 0.1in}|@{\hskip 0.1in}c@{\hskip 0.1in}|@{\hskip 0.1in}c@{\hskip 0.1in}|}
\hline
\textbf{Source} & $\boldsymbol{f}$ \textbf{(Hz)} & \textbf{Harmonics} & \textbf{LFS} \textbf{(Hz)} & \textbf{HFS} \textbf{(Hz)} & \textbf{IFO}\\
\hline
Power Mains & 60.0 & 5 & 1.0 & 1.0 & L,H\\
\hline
Violin Mode & 1373.75 & 1 & 0.1 & 0.1 & H\\
\hline
Violin Mode & 1374.44 & 1 & 0.1 & 0.1 & H\\
\hline
Violin Mode & 1377.14 & 1 & 0.1 & 0.1 & H\\
\hline
Violin Mode & 1378.75 & 1 & 0.1 & 0.1 & H\\
\hline
Violin Mode & 1379.52 & 1 & 0.1 & 0.1 & H\\
\hline
Violin Mode & 1389.06 & 1 & 0.06 & 0.06 & H\\
\hline
Violin Mode & 1389.82 & 1 & 0.07 & 0.07 & H\\
\hline
Violin Mode & 1391.5 & 1 & 0.2 & 0.2 & H\\
\hline
Violin Mode & 1372.925 & 1 & 0.075 & 0.075 & L\\
\hline
Violin Mode & 1374.7 & 1 & 0.1 & 0.1 & L\\
\hline
Violin Mode & 1375.2 & 1 & 0.1 & 0.1 & L\\
\hline
Violin Mode & 1378.39 & 1 & 0.1 & 0.1 & L\\
\hline
Violin Mode & 1387.4 & 1 & 0.05 & 0.05 & L\\
\hline
Violin Mode & 1388.5 & 1 & 0.3 & 0.3 & L\\
\hline
\end{tabular}
\egroup
\end{center}
\caption{Instrumental lines identified and cleaned before the \EatH analysis. The different columns represent: (I) the source of the line; (II) the central frequency of the instrumental line; (III) the number of harmonics in the signal-frequency range, i.e. between 1249.7 Hz and 1499.7 Hz; (IV) Low-Frequency-Side (LFS) of the knockout band; (V) High-Frequency-Side (HFS) of the knockout band; (VI) the interferometer where the instrumental lines were identified. Note that when there are higher harmonics present, the knockout bandwidth remains constant.}
\label{table:lines}
\end{table}
\pagebreak
\subsection{Signal-frequency ranges and Data Quality}
\label{section:ranges}
\begin{table}[H]
\begin{center}
\bgroup
\def\arraystretch{1.1}
\resizebox{\columnwidth}{!}{%
\begin{tabular}{|c|c|c|c|c|c|}
\hline
\hline
\textbf{Source} & \textbf{Mixed (Isolated)} & \textbf{Mixed (Left)} & \textbf{All Fake Data} & \textbf{Mixed (Right)} & \textbf{IFO}\\
\hline
\hline
Power Mains & -- -- & 1258.7976 -- 1259.2024 & 1259.2024 -- 1260.7974 & 1260.7974 -- 1261.2026 & H,L\\
Power Mains & -- -- & 1318.7915 -- 1319.2085 & 1319.2085 -- 1320.7913 & 1320.7913 -- 1321.2087& H,L\\
Violin Mode & 1372.6360 -- 1373.2140 & -- -- & -- -- & -- -- & L\\
Violin Mode & 1373.4359 -- 1374.0641 & -- -- & -- -- & -- -- & H\\
Violin Mode & 1374.1259 -- 1375.5142 & -- -- & -- -- & -- -- & H,L\\
Violin Mode & 1376.8256 -- 1377.4554 & -- -- & -- -- & -- -- & H\\
Violin Mode & 1378.0755 -- 1379.0646 & -- -- & -- -- & -- -- & H,L\\
Violin Mode & 1379.2054 -- 1379.8347 & -- -- & -- -- & -- -- & H\\
Power Mains & -- -- & 1378.7854 -- 1379.2146 & 1379.2146 -- 1380.7852 & 1380.7852 -- 1381.2148 & H,L\\
Violin Mode & 1387.1346 -- 1387.6655 & -- -- & -- -- & -- -- & L\\
Violin Mode & -- -- & 1387.9845 -- 1388.4155 & 1388.4155 -- 1388.5844 & 1388.5844 -- 1389.0156 & H,L\\
Violin Mode & 1388.7844 -- 1389.3356 & -- -- & -- -- & -- -- & H,L\\
Violin Mode & 1389.5343 -- 1390.1057 & -- -- & -- -- & -- -- & H\\
Violin Mode & 1391.0842 -- 1391.9159 & -- -- & -- -- & -- -- & H,L\\
Power Mains & -- -- & 1438.7793 -- 1439.2207 & 1439.2207 -- 1440.7791 & 1440.7791 -- 1441.2209 & H,L\\
Power Mains & -- -- & 1498.7732 -- 1499.2268 & 1499.2268 -- 1499.7170 & -- -- & H,L\\
\hline
\hline
\end{tabular}
}
\egroup
\end{center}
\caption{Signal-frequency ranges where the results might have contributions from fake data. When the results are entirely due to artificial data, the band is listed in the ``All Fake Data'' column; bands where the results comprise of contributions from both fake and real data are listed in the other three columns. The ``Mixed (Left)'' and ``Mixed (Right)'' columns are populated only when there is a matching ``All Fake Data'' entry, which highlights the same physical cause for the fake data, i.e. the cleaning. The ``Mixed (Isolated)'' column lists isolated ranges of mixed data. The list of input data frequencies where the data was substituted with artificial noise are given in table \ref{table:lines}.}
\label{table:ranges}
\end{table}
\subsection{Omitted 50$\,$mHz bands from Signal-frequency}
\label{section:excluded}
\begin{table}[H]
\begin{center}
\bgroup
\def\arraystretch{1.1}
\begin{tabular}{|l|l|l|l|l|l|}
\hline
\hline
$\boldsymbol{f_\mathsf{start}}$ \textbf{(in Hz)} & $\boldsymbol{f_\mathsf{end}}$ \textbf{(in Hz)} & \textbf{Type} & $\boldsymbol{f_\mathsf{start}}$ \textbf{(in Hz)} & $\boldsymbol{f_\mathsf{end}}$ \textbf{(in Hz)} & \textbf{Type}\\
\hline
\hline
1258.617 & 1258.717 & D & 1259.217 & 1260.717 & C\\
\hline
1291.017 & 1291.067 & D & 1292.567 & 1292.867 & D\\
\hline
1293.267 & 1293.567 & D & 1293.917 & 1294.217 & D\\
\hline
1296.367 & 1296.817 & D & 1297.517 & 1297.717 & D\\
\hline
1298.667 & 1298.967 & D & 1313.467 & 1313.517 & D\\
\hline
1318.567 & 1318.667 & D & 1319.217 & 1320.717 & C\\
\hline
1372.867 & 1373.167 & D & 1376.417 & 1376.817 & D\\
\hline
1378.517 & 1378.617 & D & 1379.217 & 1380.717 & C\\
\hline
1382.567 & -- & D & 1387.317 & -- & D\\
\hline
1387.767 & 1388.217 & D & 1388.417 & 1388.517 & C\\
\hline
1389.467 & -- & D & 1389.767 & 1390.217 & D\\
\hline
1390.467 & 1390.867 & D & 1390.967 & 1391.117 & D\\
\hline
1395.217 & 1395.467 & D & 1398.417 & 1398.667 & D\\
\hline
1399.967 & 1400.867 & D & 1400.967 & 1401.267 & D\\
\hline
1438.417 & 1438.517 & D & 1439.267 & 1440.717 & C\\
\hline
1453.467 & 1453.517 & D & 1454.967 & 1455.067 & D\\
\hline
1498.317 & 1498.467 & D & 1499.267 & 1499.667 & C\\
\hline
\hline
\end{tabular}
\egroup
\end{center}
\caption{50$\,$mHz search-frequency bands that were identified as ``disturbed'' based on Visual Inspection (Type D), or where the results were produced from ``All Fake Data'', as detailed in table \ref{table:ranges} (Type C). Both sets of bands (Type D and C) were excluded from the analysis. The first two columns list the starting frequency of the first and last 50$\,$mHz band in the contiguous range of excluded bands.}
\label{table:excluded}
\end{table}
\pagebreak
\begin{multicols}{2}
\bibliographystyle{plainnat}
\bibliography{Reference}
\end{multicols}
\end{document}